\input harvmac
%\draftmode
\let\includefigures=\iftrue
\let\useblackboard=\iftrue
\newfam\black

%Figure Stuff
\includefigures
\message{If you do not have epsf.tex (to include figures),}
\message{change the option at the top of the tex file.}
\input epsf
\def\figin{\epsfcheck\figin}\def\figins{\epsfcheck\figins}
\def\epsfcheck{\ifx\epsfbox\UnDeFiNeD
\message{(NO epsf.tex, FIGURES WILL BE IGNORED)}
\gdef\figin##1{\vskip2in}\gdef\figins##1{\hskip.5in}% blank space instead
\else\message{(FIGURES WILL BE INCLUDED)}%
\gdef\figin##1{##1}\gdef\figinbs##1{##1}\fi}
\def\DefWarn#1{}
\def\figinsert{\goodbreak\midinsert}
\def\ifig#1#2#3{\DefWarn#1\xdef#1{fig.~\the\figno}
\writedef{#1\leftbracket fig.\noexpand~\the\figno}%
\figinsert\figin{\centerline{#3}}\medskip\centerline{\vbox{
\baselineskip12pt\advance\hsize by -1truein
\noindent\footnotefont{\bf Fig.~\the\figno:} #2}}
%\bigskip
\endinsert\global\advance\figno by1}
%%%
\else
\def\ifig#1#2#3{\xdef#1{fig.~\the\figno}
\writedef{#1\leftbracket fig.\noexpand~\the\figno}%
%\figinsert\figin{\centerline{#3}}\medskip
%\centerline{\vbox{\baselineskip12pt
%\advance\hsize by -1truein\noindent
%\footnotefont{\bf Fig.~\the\figno:} #2}}
%\bigskip\endinsert
\global\advance\figno by1} \fi

\def\id{{1 \kern-.28em {\rm l}}}

\def\K3{{\bf K3}}
\def\journal#1&#2(#3){\unskip, \sl #1\ \bf #2 \rm(19#3) }
\def\andjournal#1&#2(#3){\sl #1~\bf #2 \rm (19#3) }

\def\bar{\overline}

\def\ie{{\it i.e.}}
\def\eg{{\it e.g.}}

\def\tilde{\widetilde}

\def\frac#1#2{{#1\over#2}}

\def\half{\frac12}

\def\inbar{\,\vrule height1.5ex width.4pt depth0pt}
\def\IC{\relax\hbox{$\inbar\kern-.3em{\rm C}$}}
\def\IR{\relax{\rm I\kern-.18em R}}
\def\IZ{\relax{\rm I\kern-.18em Z}}

%
%%%%%%%%%%%%%%%%%%%%%%%%%%%%%%%%%%%%
%

%
\catcode`\@=11
\def\slash#1{\mathord{\mathpalette\c@ncel{#1}}}
\overfullrule=0pt

\def\DD{{\cal D}}

\def\NN{{\cal N}}
\def\OO{{\cal O}}
\def\PP{{\cal P}}

\def\SS{{\cal S}}

\def\underrel#1\over#2{\mathrel{\mathop{\kern\z@#1}\limits_{#2}}}

\catcode`\@=12

%%%%%%%%%%%%%%%%%%%%%%%%%%%%%%%%%%%%%%%%%%%%%%%%%%%%%%%%%%%%%%

%

\def\det{{\rm det}}
\def\tr{{\rm tr}}

\def\det{{\rm det}}
\def\exp{{\rm exp}}

%%%%%%%%%%%%%%%%%%%%%%%%%%%%%%%%%%%%%%%%%%%%%%%%%%%%%%%%%%%%%%
% new defs:

\def\ie{{\it i.e.}}
\def\eg{{\it e.g.}}

\def\psibar{{\bar\psi}}

\def\rx{{\rm x}}

\def\p{\partial}
\def\ra{{\rightarrow}}

%\KaplanKR
\lref\KaplanKR{
  D.~B.~Kaplan, J.~-W.~Lee, D.~T.~Son, M.~A.~Stephanov,
  ``Conformality Lost,''
Phys.\ Rev.\  {\bf D80}, 125005 (2009).
[arXiv:0905.4752 [hep-th]].
%%CITATION = arXiv:0905.4752%%
}

%\ReyZZ
\lref\ReyZZ{
  S.~-J.~Rey,
  ``String theory on thin semiconductors: Holographic realization of Fermi points and surfaces,''
Prog.\ Theor.\ Phys.\ Suppl.\  {\bf 177}, 128-142 (2009).
[arXiv:0911.5295 [hep-th]].
%%CITATION = arXiv:0911.5295%%
}

%\RosensteinNM
\lref\RosensteinNM{
  B.~Rosenstein, B.~Warr, S.~H.~Park,
  ``Dynamical symmetry breaking in four Fermi interaction models,''
Phys.\ Rept.\  {\bf 205}, 59-108 (1991).
%%CITATION = SLAC-PUB-5349%%
}

%\KarchGX
\lref\KarchGX{
  A.~Karch and L.~Randall,
  ``Open and closed string interpretation of SUSY CFT's on branes with
  boundaries,''
  JHEP {\bf 0106}, 063 (2001)
  [arXiv:hep-th/0105132].
  %%CITATION = JHEPA,0106,063;%%
}

%\DeWolfePQ
\lref\DeWolfePQ{
  O.~DeWolfe, D.~Z.~Freedman, H.~Ooguri,
  ``Holography and defect conformal field theories,''
Phys.\ Rev.\  {\bf D66}, 025009 (2002).
[hep-th/0111135].
%%CITATION = hep-th/0111135%%
}

%\JensenGA
\lref\JensenGA{
  K.~Jensen, A.~Karch, D.~T.~Son, E.~G.~Thompson,
  ``Holographic Berezinskii-Kosterlitz-Thouless Transitions,''
Phys.\ Rev.\ Lett.\  {\bf 105}, 041601 (2010).
[arXiv:1002.3159 [hep-th]].
%%CITATION = arXiv:1002.3159%%
}

%\MateosNU
\lref\MateosNU{
  D.~Mateos, R.~C.~Myers, R.~M.~Thomson,
  ``Holographic phase transitions with fundamental matter,''
Phys.\ Rev.\ Lett.\  {\bf 97}, 091601 (2006).
[hep-th/0605046].
%%CITATION = hep-th/0605046%%
}

%\MateosVN
\lref\MateosVN{
  D.~Mateos, R.~C.~Myers, R.~M.~Thomson,
  ``Thermodynamics of the brane,''
JHEP {\bf 0705}, 067 (2007).
[hep-th/0701132].
%%CITATION = hep-th/0701132%%
}

%\MiranskyPD
\lref\MiranskyPD{
  V.~A.~Miransky and K.~Yamawaki,
  ``Conformal phase transition in gauge theories,''
  Phys.\ Rev.\  D {\bf 55}, 5051 (1997)
  [Erratum-ibid.\  D {\bf 56}, 3768 (1997)]
  [arXiv:hep-th/9611142].
  %%CITATION = PHRVA,D55,5051;%%
}

%\YamawakiVB
\lref\YamawakiVB{
  K.~Yamawaki,
  ``Quest for the Dynamical Origin of Mass: An LHC perspective from Sakata,
  Nambu and Maskawa,''
  Prog.\ Theor.\ Phys.\ Suppl.\  {\bf 180}, 1 (2010)
  [arXiv:0907.5277 [hep-ph]].
  %%CITATION = PTPSA,180,1;%%
}

%\DavisNV
\lref\DavisNV{
  J.~L.~Davis, P.~Kraus and A.~Shah,
  ``Gravity Dual of a Quantum Hall Plateau Transition,''
  JHEP {\bf 0811}, 020 (2008)
  [arXiv:0809.1876 [hep-th]].
  %%CITATION = JHEPA,0811,020;%%
}

%\CohenSQ
\lref\CohenSQ{
  A.~G.~Cohen, H.~Georgi,
  ``Walking Beyond The Rainbow,''
Nucl.\ Phys.\  {\bf B314}, 7 (1989).
%%CITATION = HUTP-88/A007%%
}

%\AppelquistWR
\lref\AppelquistWR{
  T.~Appelquist, U.~Mahanta, D.~Nash, L.~C.~R.~Wijewardhana,
  ``Gauge invariance of fermion masses in extended technicolor theories,''
Phys.\ Rev.\  {\bf D43}, 646-650 (1991).
%%CITATION = YCTP-P19-90%%
}

%\PeskinEV
\lref\PeskinEV{
  M.~E.~Peskin, D.~V.~Schroeder,
  ``An Introduction to quantum field theory,''
Reading, USA: Addison-Wesley (1995) 842 p.
}

%\AharonyAN
\lref\AharonyAN{
  O.~Aharony and D.~Kutasov,
  ``Holographic Duals of Long Open Strings,''
  Phys.\ Rev.\  D {\bf 78}, 026005 (2008)
  [arXiv:0803.3547 [hep-th]].
  %%CITATION = PHRVA,D78,026005;%%
}

%\AppelquistGY
\lref\AppelquistGY{
  T.~Appelquist, Y.~Bai,
  ``A Light Dilaton in Walking Gauge Theories,''
Phys.\ Rev.\  {\bf D82}, 071701 (2010).
[arXiv:1006.4375 [hep-ph]].
%%CITATION = arXiv:1006.4375%%
}

%\HashimotoNW
\lref\HashimotoNW{
  M.~Hashimoto, K.~Yamawaki,
  ``Techni-dilaton at Conformal Edge,''
Phys.\ Rev.\  {\bf D83}, 015008 (2011).
[arXiv:1009.5482 [hep-ph]].
%%CITATION = arXiv:1009.5482%%
}

%\SanninoQE
\lref\SanninoQE{
  F.~Sannino, J.~Schechter,
  ``Chiral phase transition for SU(N) gauge theories via an effective Lagrangian approach,''
Phys.\ Rev.\  {\bf D60}, 056004 (1999).
[hep-ph/9903359].
%%CITATION = hep-ph/9903359%%
}

%\DietrichJN
\lref\DietrichJN{
  D.~D.~Dietrich, F.~Sannino, K.~Tuominen,
  ``Light composite Higgs from higher representations versus electroweak precision measurements: Predictions for CERN LHC,''
Phys.\ Rev.\  {\bf D72}, 055001 (2005).
[hep-ph/0505059].
%%CITATION = hep-ph/0505059%%
}

%\DietrichCM
\lref\DietrichCM{
  D.~D.~Dietrich, F.~Sannino,
  ``Conformal window of SU(N) gauge theories with fermions in higher dimensional representations,''
Phys.\ Rev.\  {\bf D75}, 085018 (2007).
[hep-ph/0611341].
%%CITATION = hep-ph/0611341%%
}

%\MyersME
\lref\MyersME{
  R.~C.~Myers and M.~C.~Wapler,
  ``Transport Properties of Holographic Defects,''
  JHEP {\bf 0812}, 115 (2008)
  [arXiv:0811.0480 [hep-th]].
  %%CITATION = JHEPA,0812,115;%%
}

%\AlanenCN
\lref\AlanenCN{
  J.~Alanen, E.~Keski-Vakkuri, P.~Kraus and V.~Suur-Uski,
  ``AC Transport at Holographic Quantum Hall Transitions,''
  JHEP {\bf 0911}, 014 (2009)
  [arXiv:0905.4538 [hep-th]].
  %%CITATION = JHEPA,0911,014;%%
}

%\BergmanGM
\lref\BergmanGM{
  O.~Bergman, N.~Jokela, G.~Lifschytz and M.~Lippert,
  ``Quantum Hall Effect in a Holographic Model,''
  JHEP {\bf 1010}, 063 (2010)
  [arXiv:1003.4965 [hep-th]].
  %%CITATION = JHEPA,1010,063;%%
}

%\WaplerTR
\lref\WaplerTR{
  M.~C.~Wapler,
  ``Holographic Experiments on Defects,''
  Int.\ J.\ Mod.\ Phys.\  A {\bf 25}, 4397 (2010)
  [arXiv:0909.1698 [hep-th]].
  %%CITATION = IMPAE,A25,4397;%%
}

%\JensenGA
\lref\JensenGA{
  K.~Jensen, A.~Karch, D.~T.~Son and E.~G.~Thompson,
  ``Holographic Berezinskii-Kosterlitz-Thouless Transitions,''
  Phys.\ Rev.\ Lett.\  {\bf 105}, 041601 (2010)
  [arXiv:1002.3159 [hep-th]].
  %%CITATION = PRLTA,105,041601;%%
}

%\IqbalEH
\lref\IqbalEH{
  N.~Iqbal, H.~Liu, M.~Mezei and Q.~Si,
  ``Quantum phase transitions in holographic models of magnetism and
  superconductors,''
  Phys.\ Rev.\  D {\bf 82}, 045002 (2010)
  [arXiv:1003.0010 [hep-th]].
  %%CITATION = PHRVA,D82,045002;%%
}

%\KlebanovTB
\lref\KlebanovTB{
  I.~R.~Klebanov and E.~Witten,
  ``AdS / CFT correspondence and symmetry breaking,''
  Nucl.\ Phys.\  B {\bf 556}, 89 (1999)
  [arXiv:hep-th/9905104].
  %%CITATION = NUPHA,B556,89;%%
}

%\EvansHI
\lref\EvansHI{
  N.~Evans, A.~Gebauer, K.~Y.~Kim and M.~Magou,
  ``Phase diagram of the D3/D5 system in a magnetic field and a BKT
  transition,''
  Phys.\ Lett.\  B {\bf 698}, 91 (2011)
  [arXiv:1003.2694 [hep-th]].
  %%CITATION = PHLTA,B698,91;%%
}

%\PalGJ
\lref\PalGJ{
  S.~S.~Pal,
  ``Quantum phase transition in a Dp-Dq system,''
  Phys.\ Rev.\  D {\bf 82}, 086013 (2010)
  [arXiv:1006.2444 [hep-th]].
  %%CITATION = PHRVA,D82,086013;%%
}

%\JensenVX
\lref\JensenVX{
  K.~Jensen,
  ``More Holographic Berezinskii-Kosterlitz-Thouless Transitions,''
  Phys.\ Rev.\  D {\bf 82}, 046005 (2010)
  [arXiv:1006.3066 [hep-th]].
  %%CITATION = PHRVA,D82,046005;%%
}

%\KarchSH
\lref\KarchSH{
  A.~Karch and E.~Katz,
  ``Adding flavor to AdS / CFT,''
  JHEP {\bf 0206}, 043 (2002)
  [arXiv:hep-th/0205236].
  %%CITATION = JHEPA,0206,043;%%
}

%\BreitenlohnerBM
\lref\BreitenlohnerBM{
  P.~Breitenlohner and D.~Z.~Freedman,
  ``Positive Energy in anti-De Sitter Backgrounds and Gauged Extended
  Supergravity,''
  Phys.\ Lett.\  B {\bf 115}, 197 (1982).
  %%CITATION = PHLTA,B115,197;%%
}

%\AdamsJB
\lref\AdamsJB{
  A.~Adams and E.~Silverstein,
  ``Closed string tachyons, AdS / CFT, and large N QCD,''
  Phys.\ Rev.\  D {\bf 64}, 086001 (2001)
  [arXiv:hep-th/0103220].
  %%CITATION = PHRVA,D64,086001;%%
}

%\DymarskyUH
\lref\DymarskyUH{
  A.~Dymarsky, I.~R.~Klebanov and R.~Roiban,
  ``Perturbative search for fixed lines in large N gauge theories,''
  JHEP {\bf 0508}, 011 (2005)
  [arXiv:hep-th/0505099].
  %%CITATION = JHEPA,0508,011;%%
}

%\DymarskyNC
\lref\DymarskyNC{
  A.~Dymarsky, I.~R.~Klebanov and R.~Roiban,
  ``Perturbative gauge theory and closed string tachyons,''
  JHEP {\bf 0511}, 038 (2005)
  [arXiv:hep-th/0509132].
  %%CITATION = JHEPA,0511,038;%%
}

%\PomoniDE
\lref\PomoniDE{
  E.~Pomoni and L.~Rastelli,
  ``Large N Field Theory and AdS Tachyons,''
  JHEP {\bf 0904}, 020 (2009)
  [arXiv:0805.2261 [hep-th]].
  %%CITATION = JHEPA,0904,020;%%
}

%\SusskindDQ
\lref\SusskindDQ{
  L.~Susskind, E.~Witten,
  ``The Holographic bound in anti-de Sitter space,''
[hep-th/9805114].
%%CITATION = hep-th/9805114%%
}

%\PeetWN
\lref\PeetWN{
  A.~W.~Peet and J.~Polchinski,
  ``UV / IR relations in AdS dynamics,''
  Phys.\ Rev.\  D {\bf 59}, 065011 (1999)
  [arXiv:hep-th/9809022].
  %%CITATION = PHRVA,D59,065011;%%
}

%\HongSB
\lref\HongSB{
  D.~K.~Hong and H.~U.~Yee,
  ``Holographic aspects of three dimensional QCD from string theory,''
  JHEP {\bf 1005}, 036 (2010)
  [Erratum-ibid.\  {\bf 1008}, 120 (2010)]
  [arXiv:1003.1306 [hep-th]].
  %%CITATION = JHEPA,1005,036;%%
}

%\FujitaKW
\lref\FujitaKW{
  M.~Fujita, W.~Li, S.~Ryu and T.~Takayanagi,
  ``Fractional Quantum Hall Effect via Holography: Chern-Simons, Edge States,
  and Hierarchy,''
  JHEP {\bf 0906}, 066 (2009)
  [arXiv:0901.0924 [hep-th]].
  %%CITATION = JHEPA,0906,066;%%
}

%\SakaiCN
\lref\SakaiCN{
  T.~Sakai and S.~Sugimoto,
  ``Low energy hadron physics in holographic QCD,''
  Prog.\ Theor.\ Phys.\  {\bf 113}, 843 (2005)
  [arXiv:hep-th/0412141].
  %%CITATION = PTPKA,113,843;%%
}

%\BanksNN
\lref\BanksNN{
  T.~Banks and A.~Zaks,
  ``On the Phase Structure of Vector-Like Gauge Theories with Massless
  Fermions,''
  Nucl.\ Phys.\  B {\bf 196}, 189 (1982).
  %%CITATION = NUPHA,B196,189;%%
}

%\HillAP
\lref\HillAP{
  C.~T.~Hill and E.~H.~Simmons,
  ``Strong dynamics and electroweak symmetry breaking,''
  Phys.\ Rept.\  {\bf 381}, 235 (2003)
  [Erratum-ibid.\  {\bf 390}, 553 (2004)]
  [arXiv:hep-ph/0203079].
  %%CITATION = PRPLC,381,235;%%
}

%\MyersQR
\lref\MyersQR{
  R.~C.~Myers and R.~M.~Thomson,
  ``Holographic mesons in various dimensions,''
  JHEP {\bf 0609}, 066 (2006)
  [arXiv:hep-th/0605017].
  %%CITATION = JHEPA,0609,066;%%
}

%\AharonyUB
\lref\AharonyUB{
  O.~Aharony, M.~Berkooz, D.~Kutasov, N.~Seiberg,
  ``Linear dilatons, NS five-branes and holography,''
JHEP {\bf 9810}, 004 (1998).
[hep-th/9808149].
%%CITATION = hep-th/9808149%%
}

%\GiveonZM
\lref\GiveonZM{
  A.~Giveon, D.~Kutasov, O.~Pelc,
  ``Holography for noncritical superstrings,''
JHEP {\bf 9910}, 035 (1999).
[hep-th/9907178].
%%CITATION = hep-th/9907178%%
}

%\GinspargIS
\lref\GinspargIS{
  P.~H.~Ginsparg, G.~W.~Moore,
  ``Lectures on 2-D gravity and 2-D string theory,''
Yale Univ. New Haven - YCTP-P23-92 (92,rec.Apr.93) 197 p. Los Alamos Nat. Lab. - LA-UR-92-3479 (92,rec.Apr.93) 197 p. e: LANL hep-th/9304011.
[hep-th/9304011].
%%CITATION = hep-th/9304011%%
}

%\BeisertJR
\lref\BeisertJR{
  N.~Beisert {\it et al.},
  ``Review of AdS/CFT Integrability: An Overview,''
  arXiv:1012.3982 [hep-th].
  %%CITATION = ARXIV:1012.3982;%%
}

%\BergmanRF
\lref\BergmanRF{
  O.~Bergman, N.~Jokela, G.~Lifschytz and M.~Lippert,
  ``Striped instability of a holographic Fermi-like liquid,''
  arXiv:1106.3883 [hep-th].
  %%CITATION = ARXIV:1106.3883;%%
}

%%%%%%%%%%%%%%%%%%%%%%%%%%%%%%%%%%%%%%%%%%%%%%%%%%%
\Title{}
{\vbox{\centerline{Conformal Phase Transitions}
\bigskip
\centerline{at Weak and Strong Coupling}
}}
\bigskip

\centerline{\it  David Kutasov$^1$, Jennifer Lin$^1$ and Andrei Parnachev$^{2}$}
\bigskip
\smallskip
\centerline{${}^{1}$EFI and Department of Physics, University of
Chicago} \centerline{5640 S. Ellis Av., Chicago, IL 60637, USA }
\smallskip
\centerline{${}^{2}$Institute Lorentz for Theoretical Physics, Leiden University} 
\centerline{P.O. Box 9506, Leiden 2300RA, The Netherlands}
\smallskip

\vglue .3cm

\bigskip

\let\includefigures=\iftrue
\bigskip
\noindent
$D3$ and $D7$-branes intersecting in $2+1$ dimensions give rise at low energies to $\NN=4$ supersymmetric Yang-Mills theory coupled to defect fermions in the fundamental representation. This theory undergoes a  BKT-type phase transition from a conformal phase to one in which the fermions acquire a non-zero mass when the 't Hooft coupling of the $\NN=4$ SYM exceeds a critical value. To study this transition, we continue the parameters of the model to a regime where a gravitational description is valid. We use it to calculate the masses of mesons and the phase diagram as a function of temperature and chemical potential. We also comment on the relation of our discussion to the transition from the non-abelian Coulomb phase to a confining one believed to occur in  QCD at a critical number of flavors.  

\bigskip

\Date{}

\newsec{Introduction}

Systems of intersecting branes in string theory have proven useful in studying a variety of non-trivial dynamical phenomena in quantum field theory which are difficult to address using other techniques. Holography provides an efficient tool for analyzing such systems in particular regions of the parameter space of brane configurations. For applications one is often interested in different values of the parameters,  but the gravitational description provides a qualitative picture of the dynamics.  Quantitative predictions can then be obtained by systematically including string $(\alpha')$ and quantum $(g_s)$ corrections. 

In this paper, we discuss an example of this paradigm. The system we consider is $\NN=4$ SYM with gauge group $SU(N)$, coupled to fermions in the fundamental representation of the gauge group, which are localized on a codimension one defect. In this system one can study the dynamics of fermions with a continuously tunable interaction,  governed by the 't Hooft coupling  of the $\NN=4$ theory, $\lambda=g^2N$.  As we will see, below a critical value $\lambda_c$, the theory is conformally invariant in $2+1$ dimensions; above $\lambda_c$,  the conformal symmetry is dynamically broken  and the fermions acquire a mass.

This defect QFT can be realized in string theory as the low energy theory on $N$ $D3$-branes intersecting one or more $D7$-branes along an $\IR^{2,1}$. Some aspects of the dynamics of this brane configuration were studied in  \ReyZZ, and its relation to the Quantum Hall effect was discussed in  \refs{\DavisNV\AlanenCN-\BergmanGM} (see also \refs{\MyersME\WaplerTR\HongSB-\BergmanRF}). Our main interest in this paper is the physics in the vicinity of the phase transition at $\lambda=\lambda_c$, which does not seem to have been addressed.

Similar transitions have been discussed extensively in the context of $3+1$ dimensional non-abelian gauge theories,\foot{See \eg\ \YamawakiVB\ for a recent review and further references.} where they are expected to separate the chirally symmetric non-abelian Coulomb phase from the chiral symmetry breaking confining one. E.g., in QCD with gauge group $SU(N)$ and $F$ flavors of fermions in the fundamental representation, such a transition is expected to occur at a finite value of $x=N/F$ in the limit $N,F\to\infty$. According to this scenario, for $x<x_c$ the infrared theory is conformal and preserves the chiral $SU(F)_L\times SU(F)_R$ global symmetry. For $x>x_c$ the chiral symmetry is broken, and the fermions acquire a mass $\mu$, which scales near the transition like $\mu\simeq \Lambda_{QCD}\exp(-a/\sqrt{x-x_c})$.  The precise value of $x_c$ is not known but is believed to be around $1/4$. The large hierarchy between $\mu$ and $\Lambda_{QCD}$ is of interest in the context of (walking) technicolor (see \eg\ \HillAP\ for a review).

Phase transitions such as the ones described above are sometimes referred to as Conformal Phase Transitions (CPT's)  \MiranskyPD.  The authors of \KaplanKR\ pointed out that they occur generically when  two fixed points of the RG approach each other and annihilate. This naturally leads to the above scaling of the order parameter, which we will refer to as Miransky scaling or, following \KaplanKR, as Berezinskii-Kosterlitz-Thouless (BKT) scaling. Some recent discussions of holographic models which exhibit such scaling appear in \refs{\JensenGA\IqbalEH\EvansHI\PalGJ-\JensenVX}.

In the defect QFT that we will study, the dynamically generated mass of the fermions is expected to scale like $\Lambda_{UV}\exp(-b/\sqrt{\lambda-\lambda_c})$ as $\lambda\to \lambda_c$, where $\Lambda_{UV}$ is the UV cutoff and $b$ a constant. Since $\lambda_c$ is of order one (as we will see), superficially we cannot study the vicinity of the transition using either weak or strong coupling techniques. One approach for addressing this problem is to change the parameters such that the transition occurs at weak coupling, \ie\ $\lambda_c\ll1$. Indeed, if one takes the fermions to be localized on a $\IR^{1+\epsilon,1}$ in $\IR^{3,1}$, one finds \KaplanKR\ that for small $\epsilon$,  $\lambda_c\sim \epsilon^2$. Of course, one then has to continue the results to $\epsilon=1$ to get quantitative information about the original transition. This can in principle be achieved by including higher order perturbative contributions to the dynamics. It gives rise to the celebrated $\epsilon$-expansion that plays an important role in the study of second order phase transitions in statistical mechanics, and in many other contexts. 

In this paper we will introduce a complementary approach for studying the transition, which uses holography. As we will see, one can continue the parameters of the model to a region where the transition occurs at a large value of the 't Hooft coupling. There, $\NN=4$ SYM is described by IIB supergravity on $AdS_5\times S^5$, and the fermions by probe branes propagating in this background. The study of the physics in the vicinity of the transition amounts then (at leading order in $1/N$) to analyzing the dynamics of the probe branes in $AdS_5\times S^5$. We will use this description to calculate a number of observables, such as masses of scalar and vector mesons and the phase diagram of the model as a function of temperature and chemical potential. 

In this approach, one does not have to change the dimension of the defect on which the fermions are localized. Instead, one changes the number of $3+1$ dimensional scalar fields which couple to the defect fermions via Yukawa interactions. As in the case of the $\epsilon$-expansion, one has to include corrections to the gravity picture -- in this case, $\alpha'$ corrections -- in order to study the CPT quantitatively. However, we will see that the leading approximation already provides useful qualitative, 
and perhaps even quantitative, information.  

The plan of the paper is as follows. In section 2 we introduce the $D3-D7$ brane system and its low energy field theory description. We discuss the dynamics at weak and strong coupling, and show that this theory undergoes a phase transition at a finite value of the coupling, from a conformal phase to one in which the conformal symmetry is dynamically broken and a mass is generated for the fermions. 

In section 3 we generalize the discussion to other intersecting brane systems. We tune the parameters of these systems such that the phase transition occurs either at very weak or very strong coupling, where it can be studied using weak coupling field theory techniques and holography, respectively. We show that the transition is conformal in both regimes. This can be viewed as due to the mechanism described in \KaplanKR; as one approaches the critical coupling from below, two fixed points of the RG come together, annihilate and move off into the complex plane. This ``topological'' nature protects the transition as one varies the parameters, and makes it plausible that it is conformal in the system of section 2 as well.

In sections 4, 5 we use the gravitational description to study the system near the transition. Section 4 contains an analysis of scalar and vector mesons. As usual  in holographic systems, we find that in the gravity regime the mesons are very tightly bound, \ie\ their mass is much lower than that of the constituent fermions. We also find a finite suppression of the mass of the lowest lying scalar meson relative to those of other mesons. This scalar can be thought of as a pseudo-Goldstone boson of (explicitly) broken conformal symmetry. In section 5 we discuss the phase structure of the model as a function of temperature and chemical potential. We find that at small $T$, $\mu_{ch}$ the symmetry is broken, while at large values  of these parameters it is restored. The two phases are separated by a line of first order phase transitions.  

In section 6 we discuss the conformal phase transition in QCD from our perspective.  Further comments on our results and possible extensions appear in section 7. Some technical details of the analysis are summarized in two appendices.

\newsec{The $D3/D7$ system and its low energy dynamics}

The brane system we consider consists of $N$ $D3$-branes intersecting $F$ $D7$-branes in $2+1$ dimensions. We will take $N$ to be large, and $F=1$; it is easy to generalize the discussion to larger values of $F$. The branes are taken to be oriented as follows:
\eqn\dimensions{
\matrix{
 & 0 & 1 & 2  & 3 & 4 & 5 & 6 & 7 & 8 & 9\cr
 D3 & \rx & \rx & \rx & \rx \cr
 D7 & \rx & \rx & \rx & & \rx & \rx & \rx & \rx & \rx &\cr 
}}
The low energy spectrum of this system includes two sectors. Open strings ending on the threebranes give rise to $\NN=4$ SYM with gauge group $U(N)$, living in the spacetime labeled by $(x^0,x^1,x^2,x^3)$. Strings stretched between the threebranes and the sevenbranes give charged matter fields localized at $x^3=0$. The spectrum of these fields can be determined by a worldsheet calculation, but a quick way to find it is to note that the brane system \dimensions\ is T-dual to the well studied $D4/D8$ system that plays a role in the Sakai-Sugimoto model
\SakaiCN. Hence, the only massless fields are $2+1$ dimensional fermions which transform in the fundamental representation of $U(N)$, and can be thought of as chiral (Weyl) fermions in $3+1$ dimensions, dimensionally reduced to $2+1$ dimensions. Strings with both ends on the sevenbranes give $7+1$ dimensional massless fields, which can be treated as classical. 

We will focus on the dynamics of the defect fermions due to their coupling to the $\NN=4$ SYM fields. In the rest of this section we will discuss the dynamics first at weak coupling, where they can be studied using standard field theoretic techniques, and then at strong coupling, where one can use holography.

\subsec{Weak coupling}

For small $\lambda$, the low energy dynamics of the branes is governed by the action 
\eqn\qftaction{\SS =  \SS_{N=4}+\int d^3x (i\bar{\psi}\slash\DD \psi  + g\bar{\psi}\psi\phi^9).
}
The first part is the action of $\NN=4$ SYM, 
\eqn\snfour{\SS_{N=4}=\int d^4x{\rm Tr}\left(\frac 1 2 \DD_\mu\phi^9\DD^\mu\phi^9 - \frac 1 4 F_{\mu\nu}F^{\mu\nu} + ...  \right),}
where $\phi^9$ is a scalar field which transform in the adjoint representation of $U(N)$ and parametrizes the fluctuations of the threebranes in the $x^9$ direction. 

The second term in \qftaction\ is localized on the defect (at $x^3=0$), and describes the dynamics of the fermions. Note that only the polarizations of the gauge field along the intersection couple to the fermions -- the field $A^3$ does not couple to them directly. The Yukawa interaction is related to the fact that if we displace some or all of the $D3$-branes relative to the $D7$-brane in the $x^9$ direction, thereby giving some of the eigenvalues of the $N\times N$ matrix $\phi^9$  non-zero vevs, the corresponding components of the fermions get a mass proportional to these vevs. This also shows that there are no Yukawa interactions of the fermions with the other adjoint scalars $\phi^i$, $i=4,5,6,7,8$.

The brane configuration \dimensions\ preserves an $SO(2,1)\times SO(5)$ subgroup of the ten dimensional Lorentz group. This symmetry is manifest in the action \qftaction: the first factor is the $2+1$ dimensional Lorentz symmetry of the defect, while the second is the subgroup of the $SO(6)_R$  symmetry of $\NN=4$ SYM preserved by the defect. The action \qftaction\ furthermore preserves $2+1$ dimensional conformal symmetry. In the large $N$ limit, this symmetry remains unbroken in the quantum theory up to a critical value of the coupling, $\lambda_c\simeq 1$.

A common approach to studying the long distance dynamics of fermions in quantum field theory is to analyze the infrared properties of their self-energy  $\Sigma(p)$ (see \eg\ \refs{\CohenSQ,\AppelquistWR}), which can be thought of in terms of the non-local gauge invariant observable 
\eqn\ooww{OW(x, y)=\bar\psi(x)\PP\exp\left[ig\int_x^y A\cdot dl\right]\psi(y).}
$OW$ is an Open Wilson Line (OWL) of the sort that plays a role in gauge-gravity duality (see \eg\ \AharonyAN). Its definition depends on the choice of contour going from $x$  to $y$; a natural choice is the straight line connecting the points $x$ and $y$ \AharonyAN. The vacuum expectation value of $OW$ can be thought of as a gauge invariant generalization of the fermion two-point function that is used to define the self-energy $\Sigma$. As we discuss next, it can be used as an order parameter for the conformal phase transition. 

Suppose the dynamics of the defect fermions described by \qftaction\ is conformal for $\lambda<\lambda_c$, and massive for $\lambda>\lambda_c$ (we will justify this assumption below). For $\lambda<\lambda_c$, the one point function of $OW$ is then constrained by symmetry to have the form\foot{Here and below we omit some overall numerical factors.}
\eqn\oneptow{\langle OW(x,y)\rangle\sim i\slash\partial{1\over (x-y)^{\Delta(\lambda)-1} }}
where $\Delta(\lambda)$ is the scaling dimension of $OW$. In the free theory it is given by $\Delta(0)=2$, but in general, it is expected to depend on $\lambda$. 

For $\lambda>\lambda_c$,  the conformal symmetry is broken, and the behavior \oneptow\ is modified. The leading correction can be evaluated by  applying the operator product expansion to  \ooww,
\eqn\opeow{\lim_{x\to y}OW(x,y)\sim i\slash\partial{1\over (x-y)^{\Delta(\lambda)-1} }+
{1\over (x-y)^{\Delta(\lambda)-\Delta_b}}\bar\psi\psi(x)+\cdots}
where the ellipsis stands for contributions of higher dimension operators such as $\bar\psi\DD_\mu\psi$, and $\Delta_b$ is the scaling dimension of the fermion bilinear operator $\bar\psi\psi$. We will write it as 
\eqn\fermbi{\Delta_b(\lambda)=\Delta(\bar\psi\psi)=2+\gamma(\lambda)}
where $\gamma$ is the anomalous dimension of $\bar\psi\psi$. 

Taking the vacuum expectation value of \opeow\  one finds 
\eqn\nonconfonept{\langle OW(x,y)\rangle\sim i\slash\partial{1\over (x-y)^{\Delta(\lambda)-1} }
+{\langle\bar\psi\psi\rangle\over (x-y)^{\Delta(\lambda)-\Delta_b}}+\cdots.}
In the conformal phase $\lambda<\lambda_c$, the vev $\langle\bar\psi\psi\rangle$ vanishes and one recovers \oneptow. For $\lambda$ slightly above $\lambda_c$, this vev is non-zero but very small. Hence one can treat the second term in \nonconfonept\ as a small perturbation, for sufficiently small $|x-y|$. 

Another way of breaking the conformal symmetry is to add to the action a mass term for $\psi$, $m\int d^3x\bar\psi\psi$. For small $m$, the contribution of this term to \nonconfonept\ can be included perturbatively. One finds 
\eqn\finalonept{\langle OW(x,y)\rangle\sim i\slash\partial{1\over (x-y)^{\Delta(\lambda)-1} }
+{\langle\bar\psi\psi\rangle\over (x-y)^{\Delta(\lambda)-\Delta_b}}+{m\over(x-y)^{\Delta(\lambda)-3+\Delta_b}}+ \cdots.}
To bring \finalonept\ into a more familiar form, it is useful to rewrite it in momentum space. We can parametrize the momentum space one point function as
\eqn\momsp{\langle OW(p)\rangle \sim p^{\Delta(\lambda)-4}\left(\slash p+M(p)+\cdots\right),}
where 
\eqn\solrg{M(p)=m\left(p\over\mu\right)^\gamma+
{\langle\bar\psi\psi\rangle\over p}\left(\mu\over p\right)^\gamma}
and $\mu$ is a renormalization scale. The one point function \momsp\ is closely related to the fermion two point function:
\eqn\defsig{S_F(p) ={i\over \slash p- m - \Sigma(p)}.} 
$\Sigma(p)$ can be calculated perturbatively by summing 1PI Feynman diagrams. It can be parametrized by
\eqn\formsig{\Sigma(p)=(1-A(p))\slash p+B(p)-m,}
which leads to
\eqn\defhatsig{S_F(p) ={i\over A(p)\slash p- B(p)}= {i\over A(p)} {\slash p +M(p)\over p^2-M^2(p)},} 
with
\eqn\defm{M(p) = \frac{B(p)}{A(p)}.}
At large momenta one can expand \defhatsig\ as 
\eqn\largepsig{S_F(p) = {i\over A(p)p^2} \left(\slash p +M(p)+\cdots\right),}
which has a very similar form to \momsp. $A(p)$ is related to the anomalous dimension of the fermion $\psi$ and is typically gauge dependent \refs{\CohenSQ,\AppelquistWR}; its gauge invariant  analog in \momsp\ is the factor $p^{2-\Delta(\lambda)}$. The large momentum behavior of $M(p)$ was discussed in \CohenSQ. An analog of the operator product expansion  \opeow\ for $\bar\psi(x)\psi(y)$ leads again to an expansion of the form \solrg. Thus, the leading large momentum behavior of  $M(p)$ in \largepsig\ is given by \solrg\ and  in particular is gauge invariant \AppelquistWR. 

Note that $M(p)$ \solrg\ is a solution of the second order equation 
\eqn\clegfull{\frac{d}{dp}(p^{2}M'(p)) +C(\lambda)M(p)= 0,} 
 where the function $C(\lambda)$ is related to the anomalous dimension $\gamma(\lambda)$ \fermbi, \solrg\ via 
\eqn\anomdimc{\gamma(\lambda)=-\half+ \sqrt{{1\over4}-C(\lambda)},}
and can be calculated perturbatively in $\lambda$ using standard techniques. The one loop calculation described in appendix A yields
\eqn\csmall{C(\lambda)= \frac{5\lambda}{12\pi^2}  +O(\lambda^2).}
At weak coupling $C(\lambda)$ is small, but as the coupling increases, it can potentially become of order one, and in particular approach the value $C(\lambda_c)=1/4$ at which the anomalous dimension \anomdimc\ becomes complex. As we discuss next, this is the location of the CPT. 

In order to solve the second order differential equation \clegfull\ for the self-energy $M(p)$,  we need to specify the boundary conditions. It turns out that they are
\eqn\bcs{M'(0) = 0, \qquad M(\Lambda_{UV}) = 0,}
where $\Lambda_{UV}$ is the UV cutoff of the field theory. The first of these follows from the assumption that $M(p)$ is smooth near $p=0$. The second is the requirement that the bare fermion mass term in the Lagrangian vanishes. 

There is a subtlety regarding the boundary conditions \bcs, and more generally the definition \momsp\ of $M$. We defined this quantity as the leading deviation from the conformal behavior \oneptow. Thus, eq. \clegfull\  is valid only for $M(p)\ll p$. This condition breaks down at energies on the order of the dynamically generated mass $\mu$, or explicit mass $m$. In order to impose the first boundary condition in \bcs, we need to extend the definition of $M(p)$ beyond the linear regime described by eq. \clegfull. This is discussed at weak coupling in Appendix A, and at strong coupling later in this section. Generally, since the solution for $M$ is not expected to change much for $0<p<\mu$, one can impose the infrared boundary condition at $p\sim\mu$. 

We are now ready to discuss the solution of \clegfull, \bcs. It is easy to see that as long as $\gamma$ \anomdimc\ is real  (which, assuming $C(\lambda)$ is a monotonically increasing function of $\lambda$, occurs when $\lambda\le\lambda_c$), the only solution of this (linear) equation with the right boundary conditions is $M=0$. However, for $C(\lambda)>1/4$ (\ie\  $\lambda>\lambda_c$), non-trivial solutions exist. Defining 
\eqn\kappafull{\kappa(\lambda)=C(\lambda)-{1\over4}~,}
the general solution of \clegfull\ takes the form 
\eqn\clegsoln{M(p) = A\mu\left(\frac{\mu}{p} \right)^{\frac 12}\sin\left(\sqrt{\kappa}\ln\frac{p}{\mu} + \phi \right).}
The scale $\mu$ as well as dimensionless parameters $A$ and $\phi$ are at this point free. Two of them, \eg\ $A$ and $\phi$, can be fixed by the boundary conditions \bcs. The renormalization scale $\mu$ can be chosen  to be equal to the dynamically generated scale,
\eqn\musig{\mu=M(0).}
Any other value is related to this one by a renormalization group transformation. 

Assuming that $\mu \ll \Lambda_{UV}$ and imposing the second boundary condition in \bcs\ leads to the relation  
\eqn\hierarchy{\sqrt{\kappa}\ln{\Lambda_{UV}\over\mu}+\phi=(n+1)\pi;\qquad n=0,1,2,\cdots.}
The solution \clegsoln\ with $n=0$ starts at $M=0$ at $\Lambda_{UV}$, and monotonically increases to $M\sim\mu$ at $p=\mu$. There the linear approximation $M(p)\ll p$ breaks down, but as mentioned above, $M$ remains approximately constant over the remaining momentum interval. The dynamically generated scale is given by
\eqn\bktscaling{\mu\simeq \Lambda_{UV} \exp\left(- \frac{\pi}{\sqrt{\kappa}} \right).}
To have a large hierarchy of scales between $\mu$ and $\Lambda_{UV}$, one must be near the transition, where $\kappa\ll1$. 

The solutions with $n\ge1$ have a similar structure, but $M(p)$ has in these cases $n$ nodes in the interval $0<p<\Lambda_{UV}$.  As we will see below, these solutions have higher energies then the symmetry breaking vacuum, and are not even locally stable (in the regime we will discuss). Hence, they do not seem to play an important role in the dynamics. 

To summarize, as the 't Hooft coupling $\lambda$ increases, $C(\lambda)$ increases and the dimension $\Delta_b$ of $\bar\psi\psi$ decreases. The CPT occurs when $C(\lambda_c)=1/4$. For $\lambda<\lambda_c$, $\Delta_b$ is real and  $M(p)$ vanishes; for $\lambda>\lambda_c$ it is complex, and the lowest energy solution has $M(p) \not=0$, with large momentum behavior given by \clegsoln. At $\lambda=\lambda_c$ the anomalous dimension \anomdimc\ is $\gamma(\lambda_c)=-1/2$,  the operator 
$\bar\psi\psi$ has dimension $3/2$, $(\bar\psi\psi)^2$ becomes marginal,\foot{Note that we are assuming here that $\Delta((\bar\psi\psi)^2)=2\Delta(\bar\psi\psi)$. This assumption is valid in the large $N$ limit, where our discussion takes place,  but in general this relation receives $1/N$ corrections. We will comment on such corrections in the discussion section.} and the two leading terms in the large momentum expansion of $M$ given in \solrg\ have the same momentum dependence (up to logarithmic corrections). This is the analogue of the statement in four dimensional QCD that the transition from the non Abelian Coulomb phase to the confining one is expected to occur when $\Delta(\bar\psi\psi)=2$ and the quartic fermion coupling becomes marginal \CohenSQ. 

This picture is also in agreement with the discussion of  \KaplanKR, illustrated in figure 1. For $\lambda<\lambda_c$, the theory has two fixed points that differ in the value of the four Fermi coupling. The anomalous dimensions of $\bar\psi\psi$ at the two fixed points takes the values
\eqn\anompm{\gamma_\pm(\lambda)=-\half\pm \sqrt{{1\over4}-C(\lambda)}.}
As $\lambda\to\lambda_c$ the fixed points approach each other, and above $\lambda_c$ they move off the real axis. 
The resulting transition satisfies BKT scaling \bktscaling.

\ifig\loc{Fixed point structure as a function of the 't Hooft coupling $\lambda$.}
{\epsfxsize3.3in\epsfbox{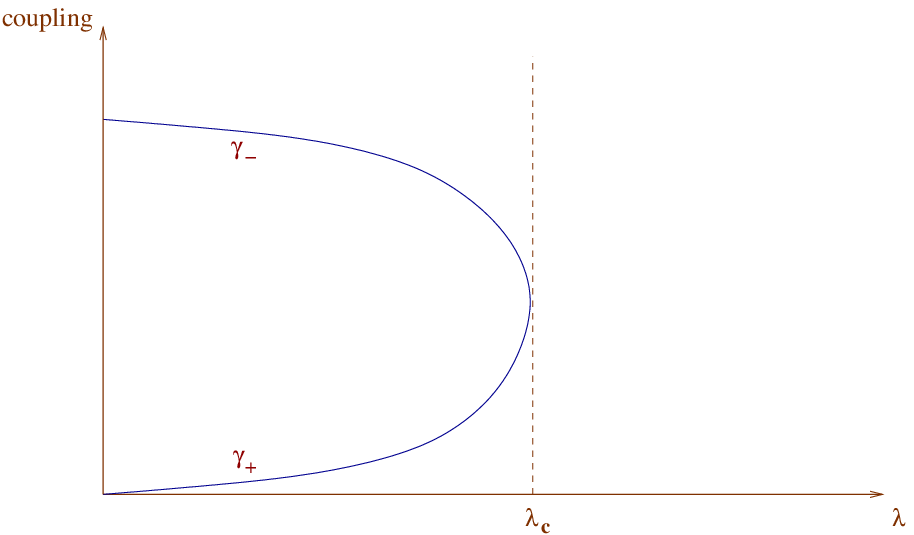}}

In the discussion above, we have assumed that the anomalous dimension of the mass operator gets to $-1/2$ at a finite value of the 't Hooft coupling, or in other words that $C(\lambda)$ gets to $1/4$ at a finite value of $\lambda=\lambda_c$. The perturbative calculation of appendix A which leads to \csmall\ is clearly not enough to establish that. However, holography enables us to analyze the system, and in particular to calculate $C(\lambda)$, in the strong coupling regime. We next turn to that calculation, which will help to determine whether a phase transition indeed takes place.

\subsec{Strong coupling}

For large $\lambda$, the dynamics of $\NN=4$ SYM is described by type IIB supergravity (more generally IIB string theory) in $AdS_5\times S^5$. The $D7$-brane \dimensions\ can be viewed as a probe propagating in this background \KarchSH; its back-reaction on the geometry can be neglected to leading order in $1/N$. To describe the intersecting brane system \dimensions\ it is convenient to write the metric on $AdS_5\times S^5$ as follows:
\eqn\dthreemetric{ds^2= \left(\frac{r}{L}\right)^2 dx_\mu dx^\mu + \left(\frac{L}{r}\right)^{2} \left(d\rho^2+\rho^2 d\Omega_4^2 +(d x^9)^2\right),}
where $\mu= 0,1,2,3,$ and $L$ is the AdS radius,\foot{We set $\alpha'=1$.} $L^4=2\pi^2\lambda$. The six dimensional space transverse to the threebranes has been split into an $\IR^5$ corresponding to the directions $(45678)$, which is described by the spherical coordinates $(\rho, \Omega_4)$, and $x^9$. Thus, these coordinates satisfy  $\rho^2 = (x^4)^2 + \cdots+ (x^8)^2$, $r^2 = \rho^2 + (x^9)^2$. The background is supported by RR five-form flux, but this will not play a role below and we will not write it explicitly.

In the brane configuration \dimensions, the $D7$-brane is stretched along the surface $x^3=x^9=0$. The induced metric on its worldvolume is given by 
\eqn\dsevenmm{ds^2= \left(\frac{\rho}{L}\right)^2 dx_a dx^a + \left(\frac{L}{\rho}\right)^{2} \left(d\rho^2+\rho^2 d\Omega_4^2\right),}
where $a=0,1,2$ are directions along the $D3/D7$ intersection. The metric \dsevenmm\ is that of $AdS_4\times S^4$; it makes manifest the conformal symmetry of the problem, as well as the other symmetries mentioned above. However, this configuration is unstable to condensation of the scalar field corresponding to the fluctuations of the $D7$-brane in $x^9$. Indeed, if we generalize the ansatz for the shape of the brane from $x^9=0$ to $x^9=f(\rho)$, we find the induced metric 
\eqn\dsevenmetric{ds^2=  \left(\frac{r}{L}\right)^2 dx_a dx^a + \left(\frac{L}{r}\right)^{2} \left([1+f'(\rho)^2] d\rho^2+\rho^2 d\Omega_4^2 \right).} 
The profile $f(\rho)$ is determined by minimizing the DBI action 
\eqn\dbia{S_{D7} =  \int d^3x \int_0^\Lambda d\rho {L^2 \over \rho^2+f(\rho)^2} \rho^4 \sqrt{1+f'(\rho)^2}}
where the prime denotes differentiation w.r.t. $\rho$ and we omitted an overall multiplicative constant.  As is usual in holography, one should think of the radial direction $\rho$ as corresponding to energy scale in the field theory. The upper limit of the integral in \dbia, $\Lambda$,  is related to the UV cutoff in the field theory via \refs{\SusskindDQ,\PeetWN}:
\eqn\cutoffmap{   \Lambda \simeq \Lambda_{UV} \sqrt{\lambda}  . }
The Euler-Lagrange equation that follows from \dsevenmetric\  is
\eqn\eom{ 
{\frac{\p}{\p\rho}}
 \left( {\frac{\rho^4}{ \rho^2+f(\rho)^2}} {\frac{f'(\rho)}{ \sqrt{1+f'(\rho)^2}}}\right)  +
{\frac{2 f(\rho)}{ (\rho^2+f(\rho)^2)^2}} \rho^4 \sqrt{1+f'(\rho)^2} = 0.   }
The ground state of the system corresponds to the lowest energy solution of this equation which satisfies the boundary conditions 
\eqn\bcsgravity{f'(0)=0,\qquad f(\Lambda)=0.}
The first is necessary for regularity of the shape of the brane near the origin of the $\IR^5$ labeled by $(45678)$; the second is the statement that we do not displace the sevenbranes from the threebranes by hand, \ie\ that the bare fermion mass vanishes.  

The simplest solution of \eom\ with these boundary conditions is $f(\rho)=0$. It preserves the scaling symmetry  $\rho \rightarrow \alpha\rho, f\rightarrow \alpha f$ of \eom, and corresponds to the conformally invariant state \dsevenmm. However, as we will see next, the ground state of this system has $f(\rho)\not=0$, and breaks the conformal symmetry.  

To find it, we start at large $\rho$, where we expect $f(\rho)$ to be small. Expanding \eom\ in $f(\rho)$ yields the linear equation of motion

\eqn\eomlim{{\frac{\p}{\p\rho}} \left( \rho^2 f'(\rho) \right) +2 f(\rho) =0. }
If we think of $f(\rho,x^a)$ as a scalar field in $AdS_4$, \eomlim\ is essentially the Klein-Gordon equation it satisfies. It is easy to check that the mass of this field is below the Breitenlohner-Freedman (BF) bound \BreitenlohnerBM. Thus, the system is unstable to its condensation.\foot{See \eg\ \refs{\AdamsJB\DymarskyUH\DymarskyNC-\PomoniDE} for other discussions of tachyons below the BF bound in AdS. In these papers the infrared instabilities due to the presence of these tachyons were associated in the dual field theory to double trace operators. In our case the tachyon is a single trace operator corresponding to a light mode of an open string whose ends lie on a $D7$-brane in $AdS_5\times S^5$ and the role of the double trace operators is played by $(\bar\psi\psi)^2$.}

The general solution of \eomlim\ is 
\eqn\realsol{
 f({\rho) =  A \; \mu \left({\frac{\mu}{\rho}}\right)^{{\frac 1 2}} 
                  \sin\left( \frac{\sqrt{7}}{2}\ln {\frac\rho\mu} +\phi \right),
                  }}
where we introduced two dimensionless parameters $A$ and $\phi$, and an arbitrary scale $\mu$. Solutions with different values of $\mu$ are related by the scaling symmetry mentioned above.
The result \realsol\ is reliable at large $\rho$; as $\rho$ decreases, the small $f(\rho)$ expansion eventually fails, and we have to go back to the full equation \eom. As in the previous subsection, we choose the scale $\mu$ to be 
\eqn\choosemu{\mu=f(0)}
and then fix the parameters $A$ and $\phi$  by imposing the boundary condition $f'(0)=0$. From \choosemu\ we find (numerically) $A\approx 0.761; \phi\approx 1.086$. To fix $f(0)$, we need to impose the second boundary condition in \bcsgravity, $f(\Lambda)=0$. This gives an infinite number of solutions, 
\eqn\soltach{\mu_n=\Lambda\exp\left(-{2\over\sqrt7}\left((n+1)\pi-\phi\right)\right),\qquad n=0,1,2,\cdots}
some of which are shown in figure 2. The $n$th solution has $n$ nodes in the interval $0< \rho < \Lambda$. 

\ifig\loc{Solutions of \eom\ with $\Lambda = 4.7305$ and $\mu = 1$. The solid, dashed and dotted lines correspond to $n=0,1,2$. The $n=1,2$ solutions have been rescaled by factors of 8 and 64, respectively.}
{\epsfxsize3.3in\epsfbox{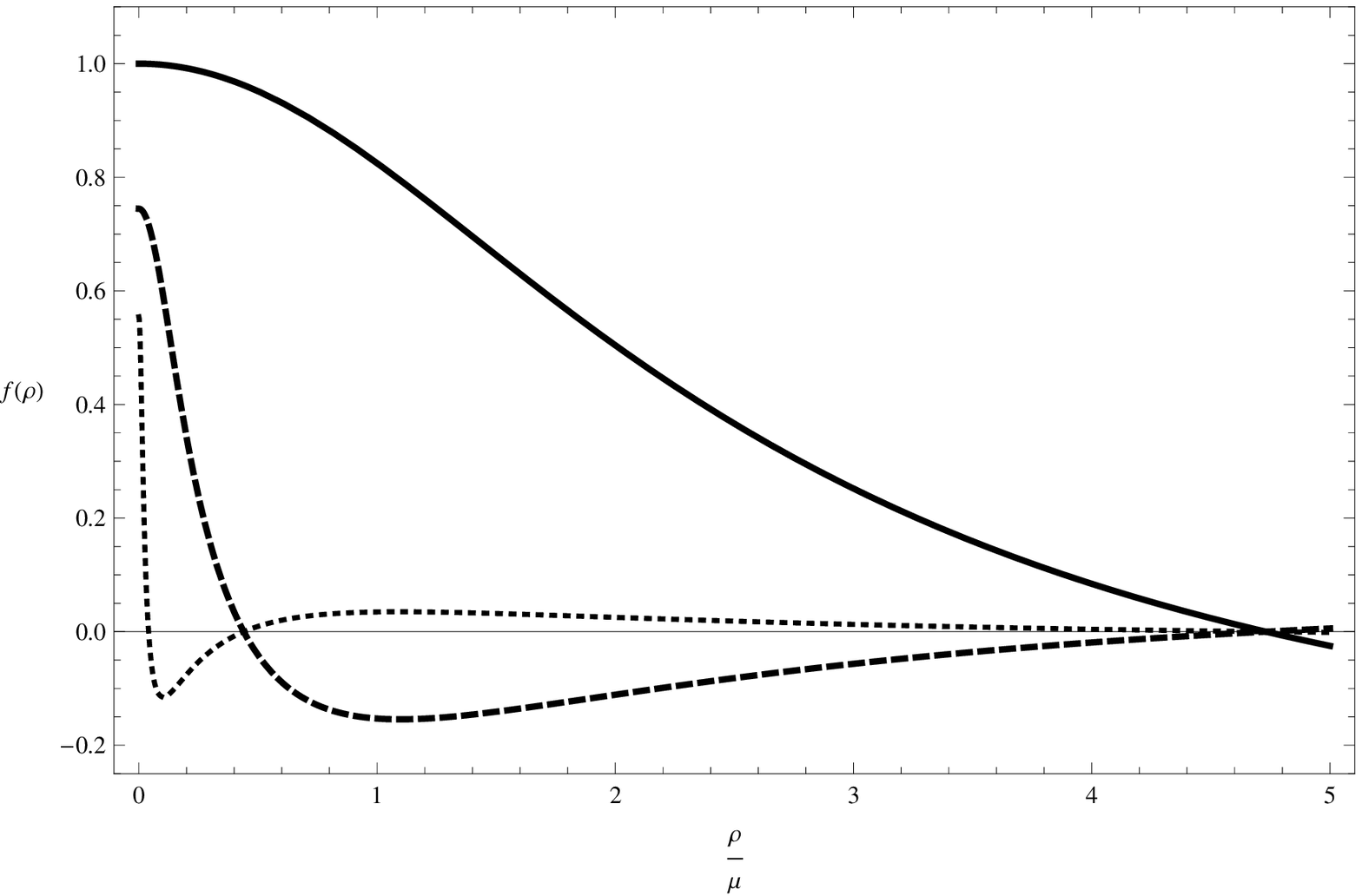}}

To identify the ground state of the system, we need to compare the energy densities of the solutions that we found, which are the values of the action \dbia\ evaluated on the solutions. 
 The energies of all the states behave at large $\Lambda$ like $L^2\Lambda^3/3$, but the differences of energies are finite as $\Lambda\to\infty$. One finds that the energy grows with $n$, so the solution with $n=0$ (no nodes) is the ground state. The conformally invariant solution $f(\rho)=0$ can be thought of as the limit as $n\to\infty$ of the others; it has the highest energy. If we set the energy of this solution to zero, the ground state has energy density
\eqn\vacen{E_0\simeq-0.17\mu^3L^2/g_s\sim-\mu^3 N/\sqrt\lambda.} 
To compute the energy density in \vacen\ we restored the factor of $1/g_s$ in the normalization of the DBI action \dbia. The $N$ dependence of the vacuum energy is compatible with the fact that we are dealing with a system of $N$ fermions, and the symmetry breaking affects their state. The  $\lambda$ dependence suggests that another scale, in addition to $\mu$, plays a role in the dynamics. As we will see later, the two important physical scales are $\mu$ and $\mu/\sqrt{\lambda}$. The former corresponds to the dynamically generated mass of the fermion, given by the energy of an open string which stretches at $\rho=0$  between the $D7$-brane (located at $x^9=\mu$) and the horizon at $x^9=0$. The latter is associated with the meson masses and the temperature of the phase transition to the symmetric phase.

From the brane perspective, one can think of the energetics as due to a repulsive force between the $D3$ and $D7$-branes, which (at strong coupling) pushes the sevenbranes away from the origin in $x^9$. This breaks the $x^9\rightarrow -x^9$ parity symmetry, as well as conformal  symmetry, and generates a mass for the fermions. The analysis of the previous subsection shows that for small $\lambda$ this repulsive force is too weak to generate a mass. 

Our treatments of the fermion self-energy $M(p)$ in the previous subsection, and of the sevenbrane profile $f(\rho)$ in this one, were very similar. Indeed, the linear equation \clegfull\ is a direct analog of \eomlim, with $M\leftrightarrow f$ and $p\leftrightarrow \rho$. The boundary conditions \bcs\ and \bcsgravity\ are the same as well. The large momentum expansion of $M$ \solrg\ is mapped to the standard large $\rho$ expansion of a scalar field in anti de-Sitter space, with the leading, non-normalizable, term corresponding to a coupling in the Lagrangian of the dual field theory, and the first subleading correction encoding the vev of the corresponding field.  

The boundary field $\bar\psi\psi$ whose coupling and expectation value appear in \solrg\ is related by holography to the bulk field $f(\rho)$. Hence it is natural to think of the large $\rho$ expansion of $f(\rho)$ as providing information about the (strong coupling limit of the) large $p$ expansion of $M(p)$. In particular, comparing \clegsoln\ to \realsol\ we see that  $\lim_{\lambda\to\infty}\kappa(\lambda)=7/4$, \ie\ $C(\lambda\to\infty)=2$. Assuming that $\kappa$ is a continuous function of $\lambda$, it has to pass through zero at a critical coupling $\lambda_c$. 

To summarize, we see that the system  \dimensions, \qftaction\ undergoes a phase transition at a finite coupling, which is out of reach of both the weak and strong coupling expansions. Indeed, for small $\lambda$ we found $M(p)=0$ and no dynamically generated scale, whereas for large $\lambda$ the dynamically generated scale, $\mu_0$ in \soltach, is comparable to the UV cutoff $\Lambda$. The transition could either be continuous, with the dynamically generated scale vanishing for $\lambda<\lambda_c$  (defined by $\kappa(\lambda_c)=0$), and continuously approaching zero as $\lambda\to\lambda_c$ from above, or it could be discontinuous, with the scale in the broken phase bounded from below by a value of order the UV cutoff.  In the former case the transition must be conformal, driven by the mechanism described in the previous subsection, and exhibit Miransky scaling \bktscaling. In the latter, the transition must occur at a coupling strictly below $\lambda_c$.

In order to distinguish between the two possibilities, we will describe in the next section two perturbative expansions that allow one to analyze the physics near the phase transition. By changing the parameters of the system  \qftaction, we will be able to push the transition towards weak or strong coupling, and show that in both limits the dynamically generated scale can be made arbitrarily small in its vicinity. This provides strong evidence that the phase transition in this system  is conformal.

\newsec{Perturbative expansions}

In the previous section we saw that the $D3/D7$ system undergoes a phase transition at a coupling of order one. In this section we will discuss approximation schemes that allow one to study this transition. To this end, we will replace the flavor $D7$-brane by a $Dp$-brane which shares $d$ spacetime dimensions with the $D3$-branes, and is extended in $n=p+1-d$ additional spatial directions.  We take the orientation of the branes to be
\eqn\dimensiondn{\epsfxsize2.7in\epsfbox{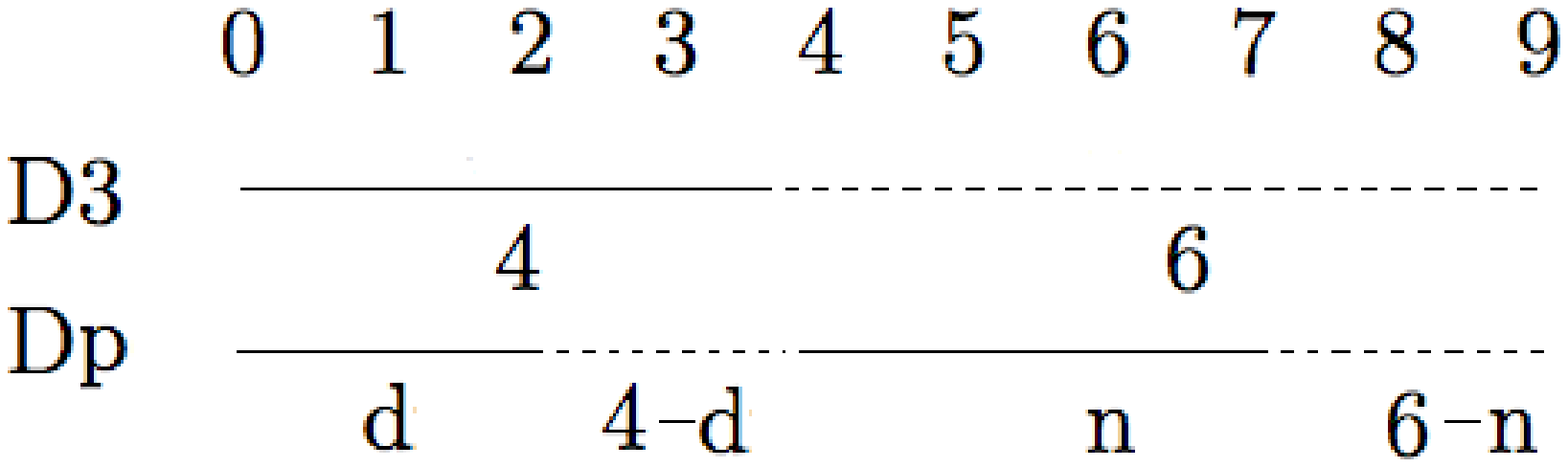}}
where solid (dashed) lines indicate directions along (transverse to) the branes. The original $D3/D7$ system \dimensions\ corresponds to $(d,n) = (3,5)$. The supersymmetric $D3/D5$ system studied in \refs{\KarchGX,\DeWolfePQ} corresponds to $d=n=3$, and one can consider many other systems with various $d$ and $n$. We will eventually take these  parameters to have general non-integer values. For $d$, this is familiar from the $\epsilon$-expansion, but we will find it useful to vary $n$ as well.  

The brane configuration \dimensiondn\ is invariant under $O(d-1,1)\times O(4-d) \times O(n) \times O(6-n)$. In addition to the $\NN=4$ SYM fields living on the $D3$-branes, the low energy spectrum contains the massless states of open  strings stretched between $D3$ and $Dp$-branes. The number of directions of space in which these strings have Dirichlet boundary conditions on one end and Neumann on the other is $\#ND= 4 +n-d$. If this number is smaller than four (\ie\ for $n<d$), the lowest lying (bosonic) excitation in this sector is tachyonic. The dynamics of open string tachyons is well understood; their presence typically means that the brane configuration can continuously decrease its energy by classical rearrangement of the branes. Our interest here is in different dynamical phenomena; therefore we will consider only $n\ge d$ where such tachyons are absent.

For $n=d$, the brane system \dimensiondn\ is supersymmetric; it preserves eight of the 32 supercharges of type IIB string theory. The massless spectrum contains in this case a hypermultiplet which transforms in the fundamental representation of $SU(N)$.  An example of this is the system considered in  \refs{\KarchGX,\DeWolfePQ}. Since our interest here is in the dynamics of defect fermions, we will restrict to the case 
\eqn\notach{n>d}
in which the lowest lying bosons in the $(3,p)$ sector are massive, with $m^2\simeq n-d$ (in string units).  The fermions can be thought of as  chiral spinors of $SO(d+5-n,1)$ decomposed under $SO(d-1,1)\times SO(6-n)$; they are singlets of $SO(4-d)\times SO(n)$. We will eventually be interested in continuing $d$, $n$ to general non-integer values, but to perform calculations at weak coupling it is convenient to take $d-n\in2 Z$ and continue the results.  

Our goal in the rest of this section is to repeat the discussion of the previous section for all $d$ and $n$ and look for regimes in which the critical  coupling $\lambda_c$ is very small or very large, so we can use perturbation theory or holography.

\subsec{Weak coupling}

The low energy effective field theory corresponding to the brane configuration \dimensiondn\ is described by the action
\eqn\qftactiontwo{\SS = \SS_{N=4}+
\int d^dx(i\bar{\psi}\slash\DD\psi  + g\bar{\psi}\Gamma_m\phi^m\psi),}
where the vector index $m$ runs over the $6-n$ transverse directions in \dimensiondn. $\phi^m$ are four dimensional fields that transform in the adjoint of $SU(N)$ and parametrize fluctuations of the threebranes in these directions. They are generalizations of the field $\phi^9$ in the previous section, and the origin of the Yukawa couplings in \qftactiontwo\ is the same as there. 

The weak coupling calculation of the fermion self-energy \defsig\ -- \largepsig\ is described in appendix A. One finds that eq. \clegfull\ generalizes to
\eqn\cleggend{\frac{d}{dp}\left(p^{d-1}M'(p)\right) +C(\lambda)p^{d-3}M(p)=0,}
where 
\eqn\formcl{C(\lambda)= \frac{\lambda}{4d\pi^2}(2(d-1)+(6-n)) + O(\lambda^2).}
The solution of \cleggend\ at finite coupling depends on the sign of 
\eqn\kappagendn{\kappa(\lambda)=C(\lambda)- \left(\frac{d-2}{2}\right)^2.} 
For $\kappa<0$ (\ie\ for sufficiently weak coupling), it takes the form
\eqn\solrggend{M(p)=m\left(p\over\mu\right)^\gamma+
{\langle\bar\psi\psi\rangle\over p^{d-2}}\left(\mu\over p\right)^\gamma,}
where
\eqn\anomdimc{\gamma(\lambda)=\gamma_+(\lambda)=-{d-2\over2}+ \sqrt{\left(\frac{d-2}{2}\right)^2-C(\lambda)}}
is the anomalous dimension of $\bar\psi\psi$. In the limit $m\to0$ the theory becomes conformal, and $M(p)$ vanishes, as in the discussion of section 2.

In this range of couplings, we also expect a second fixed point, which is formally obtained by adding to the Lagrangian the irrelevant double trace operator $(\bar\psi\psi)^2$ and flowing up the resulting RG trajectory (see figure 1). At this fixed point  the anomalous dimension $\gamma$ takes the form 
\eqn\anomdimuv{\gamma(\lambda)=\gamma_-(\lambda)=-{d-2\over2}- \sqrt{\left(\frac{d-2}{2}\right)^2-C(\lambda)}.}
Hence, the operator $(\bar\psi\psi)^2$ is relevant (for $d>2$); adding it to the Lagrangian leads back to the fixed point \qftactiontwo. Note that the dimension of $\bar\psi\psi$ at the new fixed point, $\Delta_{uv}(\bar\psi\psi)=d-1+\gamma_-$, satisfies the unitarity bound $\Delta\ge (d-2)/2$  for all $d\le 4$ (assuming $C(\lambda)>0$). This is consistent with the existence of such a fixed point for all values of the coupling. 

For $\lambda=0$ this fixed point is familiar from studies of the NJL model. As is well known (see \eg\  \RosensteinNM\ for a review), in $d<4$ dimensions this model is solvable in the large $N$ limit.  Using standard large $N$ techniques one can show that the fermion bilinear $\sigma=\bar\psi\psi$ has dimension one in the UV, in agreement with  \anomdimuv\ (with $C(\lambda=0)=0$). For $\lambda\not=0$ the model is no longer solvable, but it is natural to expect that the NJL UV fixed point can be smoothly continued to this regime.

When the coupling exceeds a critical value $\lambda_c$ for which $C(\lambda_c)=(d-2)^2/4$, the two fixed points discussed above merge and move off the real axis \KaplanKR.  The anomalous dimensions \anomdimc, \anomdimuv\ become complex, $\kappa$ turns positive, and the solution of  \cleggend\ assumes the form 
\eqn\clegsolntwo{ M(p) = A\mu\left(\frac{\mu}{p} \right)^{\frac {d-2}{2}}\sin\left(\sqrt{\kappa}\ln\frac{p}{\mu} + \phi \right).}
In this regime, conformal symmetry is dynamically broken. $A$ and $\phi$ can be calculated as in section 2, and the scale $\mu$ can be set equal to the dynamically generated scale, as in \musig. 

The transition between the two regimes occurs at the point where $\kappa$ \kappagendn\ vanishes. In section 2 we discussed the case $d=3$, where this happens at a coupling of order one. However, we see from \formcl, \kappagendn\ that if the dimension $d$ is close to two $(d=2+\epsilon)$, the critical coupling can be made small $(\lambda_c\sim \epsilon^2)$. In this case, the weak coupling approximation is valid in the vicinity of the transition. 

It is easy to understand why near two dimensions the CPT happens at weak coupling. In general, the transition occurs when the anomalous dimension \anomdimc\ takes the value $\gamma(\lambda_c)=-(d-2)/2$, so $\Delta(\bar\psi\psi)=d/2$, and $(\bar\psi\psi)^2$ is marginal. As is well known, for $d=2$ this operator is marginal already in free field theory. Thus, for $d$ slightly above two, the anomalous dimension needed to make it marginal is small. 

To summarize, we see that the CPT of section 2 can be studied using weak coupling techniques by continuing the dimension $d$ to $2+\epsilon$ and using the $\epsilon$ expansion \KaplanKR. Of course, to obtain quantitative results in $d=3$ one needs to continue to $\epsilon=1$, but already in leading order in $\epsilon$ one finds a phase transition which is qualitatively similar to what one expects; thus, the higher orders in $\epsilon$ are not expected to change the qualitative picture.  

For $0<\epsilon, \kappa\ll1$, there is a large range of momenta $\mu\ll p \ll \Lambda$ in which $M(p)$ is well approximated by the solution of \cleggend\ with $\kappa=0$ (\ie\ $C(\lambda)=(d-2)^2/4$). In this case the two terms in \solrggend\ have the same momentum dependence, and the general solution of \cleggend\ has the form 
\eqn\gensolweak{M(p)=\mu\left(\mu\over p\right)^{d-2\over2}\left(C_1\ln{p\over\mu}+C_2\right),}
where $C_1, C_2$ are dimensionless constants. From the point of view of \clegsolntwo, this is a regime in which the argument of the sine is small. Comparing coefficients, we conclude that for small $\kappa$
\eqn\Anphi{  A={C_1\over\sqrt{\kappa}},\qquad  \phi={C_2\over C_1} \sqrt{\kappa}. }
The dynamically generated scale can be obtained by imposing the boundary condition $M(\Lambda)=0$ \bcs. Looking back at \clegsolntwo, one finds the Miransky (or BKT) scaling behavior
\eqn\miransky{M(0)=\mu\simeq\Lambda \exp\left(-{\pi\over\sqrt\kappa}\right).}
Note that the fermion mass $m_\psi$ is not exactly equal to $\mu$, but is comparable to it. Indeed, $m_\psi$ is obtained by solving the equation $M(m_\psi)=m_\psi$ (see \defhatsig). The self energy $M(p)$ has the following structure: it is roughly constant and equal to $\mu$ for $p<\mu$, and then crosses over to the behavior \gensolweak\ for $p>\mu$. $p=\mu$ is in the crossover region between the two behaviors, but since both give $M(\mu)\simeq\mu$ there, it must be that $m_\psi\simeq\mu$. One can obtain the ratio $m_\psi/\mu$ more precisely by solving the equations of appendix A numerically.

\subsec{Strong coupling}

To extend the discussion of subsection 2.2 to general $d$, $n$, it is convenient to write the metric of $AdS_5\times S^5$ as
\eqn\bmetr{ ds^2 = \left(\frac r L\right)^2dx_\mu dx^\mu + \left(\frac L r\right)^2(d\rho^2 + \rho^2d\Omega_{n-1}^2 + df^2 + f^2d\Omega_{5-n}^2),}
where $\mu=0,1,2,3$, while $(\rho,\Omega_{n-1})$ and $(f,\Omega_{5-n})$ are spherical coordinates on the $\IR^n$ and $\IR^{6-n}$ transverse to the $D3$-branes in \dimensiondn. The radial coordinate of $AdS_5$ is given by $r^2 = \rho^2 + f^2$.

As before, we take the $Dp$-brane to wrap $\IR^{d-1,1}\times S^{n-1}$ and form the curve $f=f(\rho)$. 
The induced metric on the D$p$ brane is then given by
\eqn\imetr{ds^2_{Dp} = \left(\frac r L\right)^2 dx_a dx^a + \left(\frac L r \right)^2\left[(1+f'(\rho)^2)d\rho^2 + \rho^2d\Omega_{n-1}^2\right] ,}
where $a=0,1,2,\cdots, d-1$ runs over the intersection. For $f=0$, \imetr\  describes $AdS_{d+1} \times S^{n-1}$. As in the discussion of the $D3/D7$ system in section 2, this configuration is in general unstable to condensation of $f$. To study this instability, we write the DBI action for $f$, 
\eqn\gaction{S_{Dp} = \int d^dx \int d\rho \left(\frac{\rho^2+f^2}{L^2}\right)^{d-n\over2}\rho^{n-1}\sqrt{1+f'^2}.}
 The equation of motion for $f$ (the analog of \eom\ for general $d$ and $n$) is 
\eqn\geom{\frac{\partial}{\partial\rho}\left((\rho^2+f^2)^{(d-n)/2}\frac{\rho^{n-1}f'}{\sqrt{1+f'^2}}\right) + (n-d)(\rho^2+f^2)^{(d-n-2)/2}f\rho^{n-1}\sqrt{1+f'^2} = 0.}
The boundary conditions are given by \bcsgravity. The conformal invariance of the system leads to a scaling symmetry of \geom, $\rho \rightarrow \alpha\rho, f \rightarrow \alpha f$. As in the $D3/D7$ system, this symmetry is in general dynamically broken.

For large $\rho$, we expect $f$ to be small and slowly varying. In this regime, \geom\  reduces to
\eqn\glrho{ \frac{\partial}{\partial\rho} \left(\rho^{d-1}f'\right) + (n-d)\rho^{d-3}f = 0. }
This equation is identical to \cleggend\ with the map $M\leftrightarrow f$, $p\leftrightarrow\rho$, $C\leftrightarrow n-d$.  
The relation between $M$ and $f$ was discussed above (in subsection 2.2). Following that discussion, we see that the strong coupling analysis implies that 
\eqn\cinfinity{C_\infty=\lim_{\lambda\to\infty}C(\lambda)=n-d}
and \kappagendn
\eqn\kappainfinity{\kappa_\infty=\lim_{\lambda\to\infty} \kappa(\lambda)=n-d- \left(\frac{d-2}{2}\right)^2. }
When $\kappa_\infty < 0$, the system remains\foot{Assuming that $\kappa(\lambda)$ is a monotonic function.} in the conformal phase for all $\lambda$. The anomalous dimension of $\bar\psi\psi$, $\gamma(\lambda)$ \anomdimc, can be calculated at small $\lambda$ from \formcl, and for large $\lambda$ from \cinfinity; it remains real for all $\lambda$.
  
On the other hand, when $\kappa_\infty>0$, \ie\ for\foot{Note that $n_c$ is always in the range \notach.}
\eqn\ncrit{n>n_c=d+ \left(\frac{d-2}{2}\right)^2,}
the system undergoes a CPT at a coupling $\lambda_c$ for which $\kappa$ vanishes. In $d=3$ dimensions,  
\eqn\nnccrr{n_c=3+{1\over4}.}
As a check, the $D3/D7$ system (for which $n=5$) is indeed in the regime \ncrit, in agreement with the discussion of section 2. Supersymmetric systems, which have $n=d$,  are always in the conformal phase $\kappa_\infty<0$ \kappainfinity\ as one would expect.

In the regime \ncrit, the transition from conformal to massive behavior generally occurs at a coupling of order one. In order to study the transition, we vary $n$ to
\eqn\nnnccc{n=n_c+\delta}
such that the asymptotic value of $\kappa$, $\kappa_\infty=\delta\ll1$ \kappainfinity, takes a small positive value, and analyze the resulting theory in the DBI approximation.  We have thus pushed the transition into the regime where we can use holography. Of course, this analysis takes place in a theory with the wrong value of $n$; the theory we are interested in has $n=5$, while the DBI approximation is reliable for $n$ near the critical value \nnccrr. In order to obtain quantitative predictions, we need to  include higher order corrections in $\delta$. These corrections correspond  to $\alpha'$ corrections to the DBI action \gaction. 

The spirit of the approximation here is similar to that of the $\epsilon$-expansion. There, \eg\ in order to compute the critical exponents of the three dimensional Ising model which are governed by the IR fixed point of $\phi^4$ field theory in three spacetime dimensions, one continues the dimension of spacetime to $4-\epsilon$, so that the fixed point becomes perturbative. In our case, we push the interesting dynamics to strong coupling, where it can be analyzed using holography. This leads to a kind of gravitational $\epsilon$-expansion. As in the original $\epsilon$-expansion, the hope is that the physics for a particular value of $\kappa$ depends smoothly on $n$, so that the leading (in this case DBI) approximation  provides a good qualitative guide to the dynamics. 

To analyze the theory in the regime where $\kappa_\infty$ is small and positive (\ie\  $n$ is only slightly above $n_c$), we need to solve the DBI equation of motion \geom\ for $f$. At large $\rho$, this equation reduces to 
\glrho, whose solution behaves like 
\eqn\gsln{ f(\rho) = A\mu\left(\frac{\mu}{\rho} \right)^{\frac {d-2}{2}}\sin\left(\sqrt{\kappa_\infty}\ln\frac{\rho}{\mu} + \phi \right).}
The dynamically generated scale is given by 
\eqn\dyngen{f(0)=\mu\simeq \Lambda\exp\left(-{\pi\over\sqrt{\kappa_\infty}}\right),}
as in \miransky. 
As before, the value of the radial cutoff $\Lambda$ in the AdS space is related to the UV cutoff in the field theory via \cutoffmap. To focus on the physics near the transition, one can take the double scaling limit  $\Lambda\to\infty$, $\kappa_{\infty}\to 0$ (\ie\ $n\to n_c$), with $\mu$ held fixed.  We will refer to this as the BKT limit. In this limit,  \gsln\ behaves for $\rho\gg\mu$ as (compare to \gensolweak)
\eqn\solintreg{  f(\rho) = \mu \left(\mu\over\rho\right)^{d-2 \over 2} 
                                 \left(C_1 \ln{\rho\over\mu}+C_2  \right).}
The constants $C_1$ and $C_2$ can be obtained numerically. For the case of interest here, $d=3$, they are given by
\eqn\cvalues{ C_1\simeq 0.53575; \qquad  C_2\simeq 0.90421.}
As in the weak coupling analysis, one can use them to calculate $A$ and $\phi$ in \gsln, via \Anphi.  

Note that eq. \solintreg\ can be thought of as the asymptotic form of a scalar field {\it at} the BF bound. The leading term (proportional to $C_1$) can be thought of in the boundary theory as due to adding the dual operator, $\bar\psi\psi$, to the Lagrangian; it encodes the dynamically generated mass of $\psi$. $C_2$ is dual to the vev $\langle\bar\psi\psi\rangle$. We see that the breaking of conformal symmetry at the CPT is explicit, in agreement with the discussions in the technicolor literature and \KaplanKR.

So far we discussed the dynamical generation of a mass in the model \qftactiontwo. One could also add an explicit mass term to the Lagrangian, which corresponds in the brane language to separating the threebranes and sevenbranes in the transverse $\IR^{6-n}$.  At strong coupling, this is attained by changing the UV boundary conditions from $f(\Lambda)=0$ to $f(\Lambda) = f_{UV} > 0$. 
Let us define 
 $f_m = f(0)$ when $f(\Lambda) = f_{UV}$, and $f_0 = f(0)$ when $f(\Lambda) = 0$. 
The UV boundary condition leads to
\eqn\fuveq{ f_{UV}= A f_m \left( \frac{f_m}{\Lambda} 
       \right)^{\frac{d-2}{2}}\sin\left(\sqrt{\kappa_\infty}\ln \frac{\Lambda}{f_0} \frac{f_0}{f_m} + \phi \right).}
Using the fact that $\sqrt{\kappa_\infty}\ln \frac{\Lambda}{f_0}  + \phi=\pi$, we find in the BKT limit
\eqn\fuvtwo{\frac{f_{UV}}{f_m} = C_1\left(\frac{f_m}{\Lambda} \right)^{\frac{d-2}{2}}\ln\frac{f_m}{f_0} ,  }
where $C_1$ is the constant in \Anphi, \cvalues. The appearance of the UV cutoff $\Lambda$ in \fuvtwo\ can be understood as follows. We can rewrite this equation as 
\eqn\renorm{f_{UV}\Lambda^{d-2\over2}=C_1\left(f_m\right)^{d\over2}\ln\frac{f_m}{f_0}.   }
The cutoff dependence in \renorm\ implies that the coupling $f_{UV}$, which naively has dimension one, has in fact dimension $d/2$ in the quantum theory. The operator that couples to $f_{UV}$ is the mass operator $\bar\psi\psi$, and its dimension near the BKT transition is $d/2$, as discussed above. Thus, $f_{UV}\bar\psi\psi$ has dimension $d$, as necessary.

From the point of view of the holographic bulk description, in the BKT regime the scalar $f(\rho)$ which describes radial fluctuations of the $Dp$-brane in the transverse directions has a mass slightly below the BF bound. On the other side of the transition, where $n$ is slightly below $n_c$ \ncrit, the mass of $f$ is slightly above the BF bound and the $AdS$ vacuum \imetr\ with $f=0$ is stable. As is well known  \KlebanovTB, scalars whose mass is slightly above the BF bound can be quantized in two different ways, which lead to two distinct CFT's  differing by double trace operators. These CFT's were discussed in the previous subsection; they give rise to the anomalous dimensions \anomdimc, \anomdimuv.

Since for $\kappa_\infty>0$ the dynamics generates a mass for the fermions, one expects the trace of the stress tensor to be non-zero. To calculate it, we need to evaluate the action \gaction\ on the solution of the equation of motion \geom. As mentioned in section 2, this action is UV divergent, but differences of actions for different solutions are finite. Setting to zero the energy of the conformally invariant solution $f(\rho)=0$, one finds\foot{The numerical computation of the energy density in the BKT limit at strong
coupling is outlined in Appendix B.}
\eqn\vacaction{S_0\sim-{1\over g_s}L^{n-d}\mu^d\sim -N\lambda^{\left(d-2\over4\right)^2-1}\mu^d,}
where we used the fact that we are in the BKT regime $n\sim n_c$ \ncrit, and restored the factor of $1/g_s$ in the normalization of the DBI action \gaction. Eq. \vacaction\ implies that the stress-tensor has a non-zero vev:
\eqn\stresst{\langle T_{ab}\rangle= S_0\eta_{ab}.}
This is a very reasonable result; in the BKT limit, the vacuum energy can be expressed in terms of quantities that remain finite in this limit, the scale $\mu$ and 't Hooft coupling $\lambda$. This result also suggests that the theory contains extra scales, in addition to $\mu$. In the next section we will analyze the spectrum of mesons and see that this is indeed the case.

Finally, we would like to comment on the solutions with nodes, which were briefly discussed in section 2 (around eq. \soltach). For small $\kappa$, the solution with $n$ nodes satisfies
\eqn\Sigmann{f(0)=\mu_n\simeq\Lambda\exp\left(-{\pi(n+1)\over\sqrt{\kappa_\infty}}\right)\simeq\mu\exp\left(-{\pi n\over\sqrt{\kappa_\infty}}\right).}
In the BKT limit, all $\mu_n$ with $n\ge 1$ go to zero, so all these solutions approach the flat solution $f(\rho)=0$. It is also easy to calculate their energy densities. This can be done by splitting the $\rho$ integral  in \gaction\ into two parts: from $\rho=0$ to the first node, and from the first node to $\rho=\Lambda$. The first integral is identical to the one for the ground state, and gives 
\eqn\ssnn{S_n\sim -N\lambda^{\left(d-2\over4\right)^2-1}\mu_n^d.}
The second can be shown to vanish using a combination of two facts. One is that beyond the first node, the linear approximation \glrho\ is excellent for small $\kappa$, and one can replace the full action \gaction\ by a quadratic one. The second is that on-shell the integral of the quadratic Lagrangian reduces to boundary terms, which vanish since $f(\rho)=0$ at the boundaries (both of which are nodes). Thus, we see that the actions of the solutions with nodes \ssnn\ all vanish in the BKT limit, in agreement with the fact that they all approach the trivial solution.

\newsec{Spectrum of fluctuations}

In the previous section, we determined the shape of the probe brane $f(\rho)$ in the BKT regime and found that the theory dynamically generates a scale $\mu=f(0)$. In this section, we will study small excitations about this vacuum, which correspond to mesons. 
We will see that the meson masses are set by a new scale $\bar\mu$, defined as
\eqn\defmubar{    \bar\mu \equiv {\mu\over L^2} \simeq {\mu\over \sqrt{\lambda} }. }
This relation between the meson mass scale and the fermion mass scale is similar to that of
other holographic models (see \eg\ \MyersQR).

We first consider $\sigma$-mesons, which correspond to fluctuations of the order parameter for  the breaking of conformal symmetry. The lowest of these is the would-be Goldstone boson of broken scale invariance, the analog of the techni-dilaton in walking technicolor theories. We will see that it is lighter than the other states in the spectrum, but not parametrically so. We will also discuss vector mesons,  the analogs of the (techni-) $\rho$-meson in QCD (technicolor). 

Since the vacuum with $f(\rho)\not=0$ spontaneously breaks the global $O(6-n)$  symmetry to $O(5-n)$, when $n<5$ one expects to find massless Goldstone bosons that parametrize the coset $O(6-n)/O(5-n)$. We will describe them and calculate their mass when the symmetry is broken explicitly. 

\subsec{$\sigma$-mesons}

To study radial excitations of the probe brane, we expand the radial scalar $f$ \bmetr\ around the solution of \geom,  
\eqn\dbiexp{f(\rho, x^a) = f(\rho) +y(\rho, x^a).}
Since the DBI action preserves $d$ dimensional Lorentz symmetry, it suffices to take the perturbation $y$ to be a function of $\rho$ and $t$. For such configurations, the action takes the form 
\eqn\emdbi{S_{Dp} = -\int d^dx\int d\rho \frac{\rho^{n-1}}{r^{n-d+2}}\sqrt{r^4- L^4{\dot y}^2 + (f'+y')^2r^4}.}
If $y=0,$ this agrees with \gaction\ (up to an overall constant, which we do not keep track of here and below). 

Expanding \emdbi\ around the solution $f(\rho)$ and keeping only terms quadratic in $y$, we find the action
\eqn\lineardbi{S_2 =-\int d^dx \int d\rho \left(A(\rho)y^2 + B(\rho)y'^2 - L^4 C(\rho)\dot{y}^2\right) ,}
where 
\eqn\abcrho{\eqalign{A(\rho) = &\frac{d-n}{2}\frac{\rho^{n-1}((d-n-1)f^2+\rho^2)}{(f^2+\rho^2)^{2-\frac{d-n}{2}}}\sqrt{1+f'^2} + \frac{\partial}{\partial\rho}\left(\frac{(n-d)\rho^{n-1}ff'(f^2+\rho^2)^{\frac{d-n}{2}-1}}{\sqrt{1+f'^2}}\right) \cr
B(\rho) = &\frac{\rho^{n-1}(f^2+\rho^2)^{\frac{d-n}{2}}}{2(1+f'^2)^{3/2}}\cr
C(\rho) = &\frac{\rho^{n-1}(f^2+\rho^2)^{\frac{d-n}{2}-2}}{2\sqrt{1+f'^2}}.}}
To find the spectrum, we take $y(\rho,t)$ to have the form $y(\rho, t) = \psi(\rho) e^{imt}$.  The linear equation of motion for  $\psi$ that follows from  \lineardbi\ is     
\eqn\xeom{\frac{\partial}{\partial\rho}(B(\rho)\psi'(\rho)) + m^2L^4C(\rho)\psi(\rho) - A(\rho)\psi(\rho) = 0}
with the boundary conditions \bcsgravity, $\psi'(0) = \psi(\Lambda) =0.$ Since \xeom\ is a linear equation, the overall scale of the solution is arbitrary. Thus, \xeom\ is over-constrained and one expects solutions to exist for a discrete set of masses $m$. 

We now specialize to the physical case $d=3$. For large $\rho$, the equation for the fluctuations \xeom\ becomes identical to the equation of motion \glrho\ for the embedding function $f(\rho)$ in the linear regime. Hence, its solution takes the form \gsln,
\eqn\eqfluc{ \psi(\rho) \approx A_\sigma\, \mu 
\left(\mu\over\rho\right)^{1\over2} \sin\left(\sqrt{\kappa_\infty}\log{\rho\over\mu}+\phi_\sigma\right).  }
In order for $\psi(\rho)$ to vanish at the UV cutoff, the phase $\phi_\sigma$ in \eqfluc\
must be exactly equal to $\phi$ in \gsln. 

In the BKT limit, \eqfluc\ reduces to (see the discussion around \solintreg)
\eqn\solintnew{  \psi(\rho) \approx \mu \left(\mu\over\rho\right)^{1 \over 2} 
                                 \left( C_1^{(\sigma)} \ln{\rho\over\mu}  + C_2^{(\sigma)} \right).  }
The values of $ C_1^{(\sigma)}$ and $ C_2^{(\sigma)}$ can be obtained by
numerically integrating eq. \xeom\ starting from $\rho=0$, with the initial conditions 
$\psi(0)=1; \psi'(0)=0$.
Varying the value of $m^2$ leads to different values of $ C_1^{(\sigma)}$ and $ C_2^{(\sigma)}$;
the spectrum of fluctuations is obtained by demanding $\phi_\sigma=\phi$ or (see \Anphi, \cvalues)
\eqn\specdet{  C_2^{(\sigma)}/ C_1^{(\sigma)}=C_2/ C_1\simeq 1.6877.   }

\ifig\loc{$C_2^{(\sigma)}/C_1^{(\sigma)}$ as a function of $m^2/\bar\mu^2$.}
{\epsfxsize3.3in\epsfbox{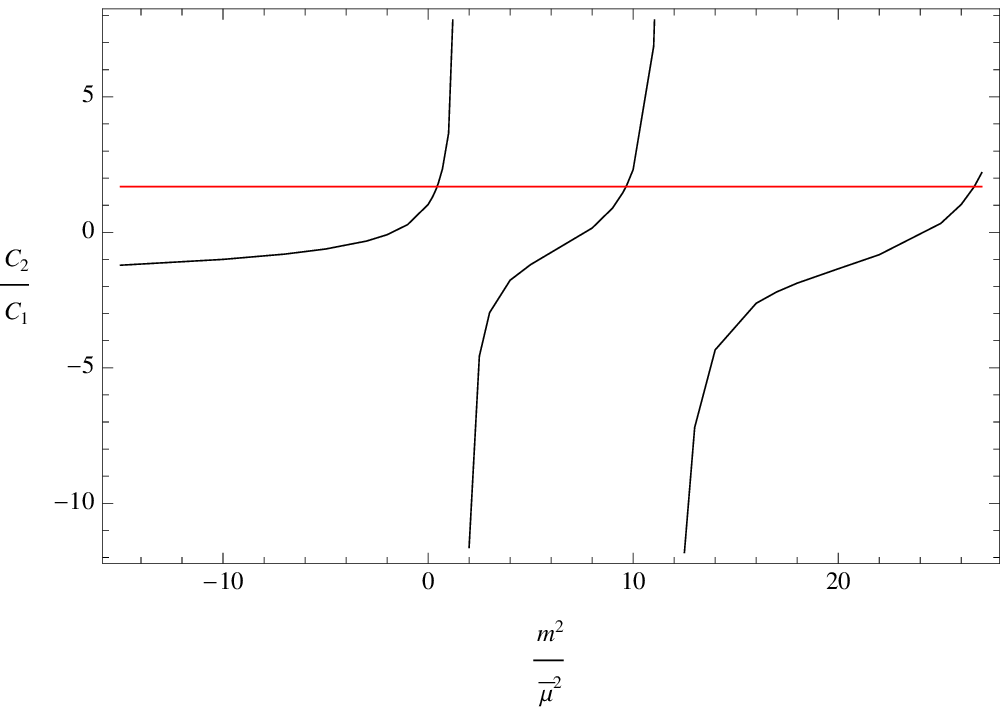}}

\noindent
In fig. 3 we plot the results of the numerical integration. Note that eq. \specdet\ is never satisfied for negative $m^2$, which is consistent with the stability of the vacuum. The lightest few states have 
\eqn\dilaton{m^2/\bar\mu^2\approx 0.44, 9.65, 26.63, 51.35, 84,\cdots,} 
where $\bar\mu$ is definied in \defmubar.
This spectrum is well described by the general formula 
\eqn\approxsigma{m_n^2/\bar\mu^2\approx 3.89n^2 + 5.32n + 0.44; \qquad n=0,1,2,\cdots.}
It exhibits the characteristic behavior $m_n\sim n$ at large excitation number $n$. This is similar to other holographic models, although here the behavior does not have an obvious Kaluza-Klein interpretation. We also see that the typical mass scale of the mesons is 
$\bar\mu \sim \mu/\sqrt\lambda$. Thus, at large $\lambda$ the mesons are deeply bound, as in \MyersQR.

Another notable fact is that the lowest lying meson, which corresponds to $n=0$ in \approxsigma, is quite light relative to the others. Its mass is $m_0\simeq 0.66\bar\mu$, while the next lightest meson has mass $m\simeq 3.1\bar\mu$, and the asymptotic separation between subsequent masses in \approxsigma\ is $m_{n+1}-m_n\simeq 1.97\bar\mu$. Thus, $m_0$ is smaller by a factor of three to five than the typical mass scales in the problem, but the separation between the two scales is not parametric in the BKT limit.  The lowest lying $\sigma$-meson can be thought of as a pseudo-NG boson of broken conformal symmetry. Since the breaking is explicit, it is massive, but is nevertheless relatively light.  

A similar issue has been discussed in the technicolor literature, where the question is whether as one approaches a CPT by tuning the number of flavors in QCD, the $\sigma$ meson is parametrically lighter than all the other mesons; see \eg\ \refs{\AppelquistGY,\HashimotoNW} for recent discussions and \refs{\SanninoQE\DietrichJN-\DietrichCM} for earlier work. There, the answer to this question is unclear, as it requires finding the spectrum of light states in a strongly coupled field theory. 

Since many of the arguments for a light dilaton, such as those of \AppelquistGY, apply to our case, it is interesting to see what happens here. Our results do not support the point of view advocated in \AppelquistGY\ that as one approaches the CPT, the ratio of mass of the dilaton to that of other mesons goes to zero. However, we did find that the mass of the lightest $\sigma$-meson is somewhat suppressed. Interestingly, a similar possibility was suggested in QCD in \HashimotoNW\ and references therein. In a certain (uncontrolled) approximation it was found that the techni-dilaton is lighter by about a factor of three than the lightest vector mesons.  We will discuss vector mesons in our model in the next subsection, and will find a similar ratio. 

So far we have discussed the radial excitations of the ground state, which from the point of view of the system with small but finite $\kappa$ (and thus finite UV cutoff $\Lambda$) is the configuration with no nodes in the interval $\rho\in (0,\Lambda)$. One can apply the same tools to analyze the solutions with nodes discussed in the previous sections. For the solution with $n$ nodes, one finds $n$ tachyons with masses of order 
\eqn\tachyons{m_i^2 \sim - \bar\mu^2 \exp\left(-\frac{2\pi (n - i)}{\sqrt{\kappa_\infty}} \right); \qquad i=1,\cdots, n.}
These tachyons signal perturbative instabilities, due to the fact that the system can continuously lower its energy by eliminating the nodes. The time scales for this process  correspond to the masses \tachyons. The fastest time scale, corresponding to $i=n$ in \tachyons, is of order $\bar\mu$, as one would expect. The time scales corresponding to $i<n$ are much longer, but involve fine-tuned dynamics. 

\subsec{Vector mesons}

To compute the spectrum of vector mesons, we turn on the gauge field on the worldvolume of the probe brane. The DBI action then reads
\eqn\dbigauge{S =- \int d^dx \int d\rho \sqrt{-\det(g + 2\pi F)},}
where $g$ is the induced metric on the brane. Expanding in the field strength $F$  and keeping only quadratic terms, we find
\eqn\dbigaugeq{S = S_0 - \int d^dx\int d\rho\left(\frac{\rho^2+f^2}{L^2}\right)^{\frac{d-n}{2}}\rho^{n-1}\sqrt{1+f'}\left[\frac {F_{a\rho}F^{a\rho}}{2(1+f'^2)} + \frac{L^4F_{ab}F^{ab}}{4(\rho^2+f^2)^2} \right],}%
where the indices along the intersection $a,b$ are raised with the flat metric $\eta^{ab}$. 

 For $F=1$ (one flavor brane) one can choose the gauge $A_\rho=0$. For larger $F$, the dynamics in general breaks the $U(F)$ global symmetry to $\prod_i U(F_i)$ with $\sum_iF_i=F$, and the off-diagonal components of $A_\rho$ play a role in the analysis of NG bosons. 

The remaining components of the gauge field can be expanded in an orthonormal basis
\eqn\gaugefields{A_a(x^b, \rho) = \sum_n B_a^{(n)}(x^b)\psi_n(\rho).} 
Suppose that the $\psi_n$ are solutions of 
\eqn\psieqn{-G^{-1}\partial_\rho(K\partial_\rho\psi_n) = m^2_n\psi_n,}
where
\eqn\gk{\eqalign{K(\rho) &= \frac{\rho^{n-1}}{(\rho^2+f^2)^{\frac{n-d}{2}}\sqrt{1+f'^2}}   \cr
G(\rho) &= \frac{L^4\rho^{n-1}\sqrt{1+f'^2}}{(\rho^2+f^2)^{\frac{n-d+4}{2}}},}}
and we impose the orthonormality condition
\eqn\psinorm{\int d\rho G(\rho)\psi_m(\rho)\psi_n(\rho) = \delta_{mn}.}
Then integration by parts leads to 
\eqn\integration{\int d\rho K \partial_\rho\psi_m\partial_\rho\psi_n = m^2_n\delta_{mn}} 
and
\eqn\dbigaugetwo{S = S_0 - \int d^dx \sum_n\left(\frac 1 4 F_{ab}^{(n)}F^{ab(n)} + \frac{m^2_n}{2}B_a^{(n)}B^{a(n)}\right).}
Hence, $m_n$ are the masses of the $d$-dimensional vector mesons. To calculate them, one needs to solve the eigenvalue problem  \psieqn\ .

Solutions of \psieqn\ behave at large $\rho$ like $\psi(\rho)\simeq c_1+c_2\rho^{-(d-2)}+\cdots$. Perturbations with $c_1\not=0$ are non-normalizable; thus we set it to zero and use the shooting method to determine the masses of the vector mesons (again, setting $d=3$ and $n \rightarrow n_c$). The first few masses are 
\eqn\vectormass{m^2/\bar\mu^2\approx   3.08, 15.12, 34.87, 62.32, 97.46, 140.31, 190.86,
249.11,\cdots.} 
They are well described by
\eqn\genvector{m_n^2/\bar\mu^2 \approx   3.85 n^2 + 0.497n- 1.27.}
We see that the spectrum of vector mesons is similar to that of the scalar mesons \approxsigma, except for the absence of an  anomalously light meson, as one would have expected. 

\subsec{Pions}

In the $D3/D7$ system of section 2, the dynamics breaks a discrete $Z_2$ symmetry $x^9 \rightarrow -x^9$. More generally, the theory has an $O(6-n)$ global symmetry, that is dynamically broken to $O(5-n)$. Unlike the breaking of conformal symmetry, the breaking of $O(6-n)$ is spontaneous. Hence, one expects to find in the spectrum massles Nambu-Goldstone bosons parametrizing the coset $O(6-n)/O(5-n)$; we will refer to them as pions. To identify them, recall that in the geometry \bmetr, the $Dp$-brane is located at a point on the $5-n$ dimensional sphere $S^{5-n}$ labeled by $\Omega_{5-n}$. The residual $O(5-n)$ symmetry is the subgroup of the isometry group of the sphere which preserves that point. 

The NG bosons correspond to slow motion of this point on the sphere. Focusing on one of them, we can parametrize the $AdS_5\times S^5$ geometry as follows:
\eqn\anglemetr{ds^2 = r^2 dx_\mu dx^\mu + r^{-2}(d\rho + \rho^2d\Omega^2_{n-1} + df^2 + f^2d\theta^2 + (f\sin\theta)^2 d\Omega^2_{4-n}) .}
The angle $\theta$ parametrizes one of the directions on the sphere. The NG boson associated with it is obtained by promoting $\theta$ to be a function of $\rho, t$. Then the DBI action of the brane generalizes to

\eqn\emdbitwo{S_{Dp} = -\int d^dx\int d\rho \frac{\rho^{n-1}}{r^{n-d+2}}\sqrt{(1+f'^2+f^2\theta'^2)r^4  - L^4f^2(1+f'^2)\dot{\theta}^2}.}
The quadratic action for $\theta$ is
\eqn\lineardbitwo{S_2 = \int d^dx\int d\rho [L^4E(\rho)\dot{\theta}^2-D(\rho)\theta'^2], }
where 
\eqn\derho{\eqalign{D(\rho) &= \frac{f^2\rho^{n-1}(f^2+\rho^2)^{\frac{d-n}{2}}}{2\sqrt{1+f'^2}} \cr
E(\rho) &=  \frac 1 2f^2\rho^{n-1}(f^2+\rho^2)^{\frac{d-n-4}{2}}\sqrt{1+f'^2}.}}
As before, we make a plane wave ansatz $\theta(\rho, t) = \tilde{\theta}(\rho)e^{im_\pi t}$. Then the eom reads
\eqn\xeomtwo{\frac{\partial}{\partial\rho}(D(\rho)\tilde{\theta}'(\rho)) + L^4m_\pi^2E(\rho)\tilde{\theta}(\rho) = 0.} 
The general solution of \xeomtwo\ behaves at large $\rho$ like $a\ln\rho+b$. Normalizable modes have $a=0$. Thus, the mode
$\tilde\theta = $ constant, which solves \xeomtwo\ with $m_\pi=0$,  gives a normalizable perturbation.

\bigskip

\ifig\loc{$m_\pi^2/\mu^2$ as a function of $f_{UV}\sqrt{\Lambda}/\mu^{3/2}$ in the BKT limit for $d=3$.} {\epsfxsize3.3in\epsfbox{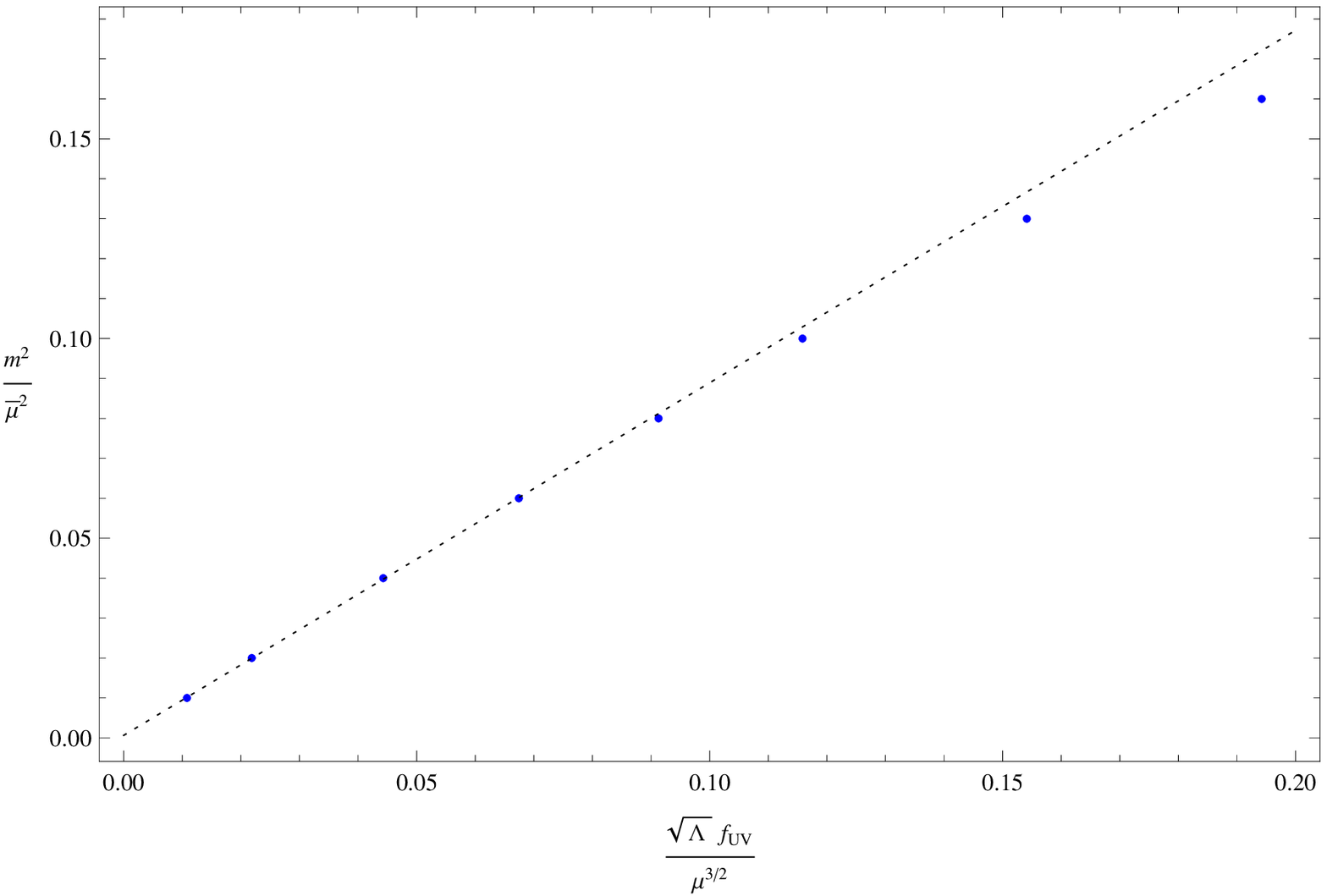}}

\bigskip

As discussed in Section 3, we can separate the threebranes and sevenbranes by changing the UV boundary condition to $f(\Lambda) = f_{UV}$, while $\theta(\Lambda)=0$; this breaks the $O(6-n)$ rotational symmetry explicitly, and gives the NG boson a mass.  The scaling of the pion mass with symmetry breaking parameters in QCD is given by the Gell-Mann-Oakes-Renner formula 
\eqn\gmor{F_\pi^2 m^2_\pi = \sigma|\langle\bar{\psi}{\psi}\rangle|,} 
where $\sigma$ is the explicit symmetry breaking mass parameter and $F_\pi$ the pion decay constant.
The analog of this in our problem can be read off fig. 4, and takes the form
\eqn\ourgmor{m^2_\pi \simeq 5.165L^4f_{UV}\mu\left(\frac{\Lambda}{\mu} \right)^{\frac{d-2}{2}}.}
As discussed around eq. \renorm,  we can identify $\sigma$ with $ f_{UV}\Lambda^{d-2\over2}$. Eq. \ourgmor\ takes then the form \gmor, with $\langle\bar{\psi}{\psi}\rangle\sim \mu^{d\over2}$, $F_\pi\sim \mu^{d-2\over2}$.

\newsec{Finite temperature and chemical potential}

In this section, we will discuss the effect of turning on finite temperature $T$ and chemical potential $\mu_{ch}$ for the fermion number. We will work at strong coupling and in the BKT limit described in section 3. As usual, to construct the phase diagram one needs to evaluate the DBI action for various types of D-brane embeddings. We will see that at sufficiently high temperature and chemical potential, the symmetric phase is thermodynamically preferred. Hence, in the $\mu_{ch}-T$ plane there is a line of first order phase transitions which separates the phase with dynamically generated mass and broken symmetry from the symmetric phase.

The characteristic scale of the temperature of this phase transition is set by the meson mass,
$\bar\mu$ (defined in \defmubar), while the scale of the chemical potential is set by the fermion mass $\mu$.
 It will be convenient to rescale the variables 
\eqn\varres{    r\ra {r\over L^2},\qquad f\ra {f\over L^2},\qquad \rho\ra {\rho\over L^2}   }
\ie\ $r_{\rm here}=r/L^2$, etc. As a result, the dynamically generated scale  is expressed as
\eqn\dynfm{    \bar\mu = f(0)\simeq \Lambda_{UV}\exp\left(-{\pi\over\sqrt{\kappa}}\right). }

\subsec{Finite temperature}

To study finite temperature physics at strong coupling, we replace $AdS_5\times S^5$ by the AdS-Schwarzschild geometry
\eqn\AdSbh{ds^2 = L^2 \left( r^2(F(r)dt^2 + dx_\alpha dx^\alpha) + \frac{dr^2}{r^2F(r)} + d\Omega_5^2 \right), }
where $\alpha=1,2,3,\, r_h = \pi T$, and 
\eqn\Fdef{ F(r) = 1 - \left(\frac{r_h}{r}\right)^4. }
The induced metric on the probe $Dp$-brane \dimensiondn\ is 
\eqn\dnbh{ ds^2_{Dp} = L^2 \left( r^2(F(r)dt^2 + dx_i^2) + \frac{1}{r^2}\left[\frac{1-F(r)}{r^2F(r)}(\rho + ff')^2 + (1+f'^2)\right] d\rho^2 + \frac{\rho^2}{r^2}d\Omega^2_{n-1} \right), }
where $i=1,2,\cdots, d-1$.  The free energy of the system is given by the DBI action,  
\eqn\ftaction{S \simeq \int d\rho\ r^{d-n}\rho^{n-1}\sqrt{F(\rho)(1+f'^2) + \frac{1-F(\rho)}{r^2}(\rho + ff')^2}.}
The asymptotic form of the action and equations of motion at energies large compared to the temperature (\ie\ $\rho\gg r_h$) are given by eqs. \gaction, \geom. The full equations of motion are complicated but can be solved numerically.  As before, we introduce the UV cutoff $\Lambda_{UV}$ and impose the Dirichlet boundary conditions $f(\Lambda_{UV})=0$.

In the presence of the black hole, the probe brane can take one of a few qualitatively different shapes. One is a ``Minkowski embedding'' in which the brane does not intersect the horizon. Regularity of the brane at $\rho=0$ leads in this case to the boundary conditions \bcsgravity. The qualitative form of the resulting solution is expected to be the following. At low temperatures $(r_h\ll\bar\mu)$, the effect of the black hole on the solution of section 3 is small, and in particular the system exhibits dynamical symmetry breaking.  As the temperature increases, the black hole both grows in size and gravitates more strongly. Eventually, one reaches a critical temperature $T_c$ at which the Minkowski solution approaches the horizon. As the temperature increases further, the intersection point of the brane and the horizon may move. 

In addition to the two types of solution described above, there is always a trivial solution, described by $f(\rho)=0$, which corresponds to the phase where mass is not generated dynamically and parity is not broken. At low temperatures, we expect the symmetric solution to have higher free energy than the Minkowski one, and thus be thermodynamically disfavored. However, above a certain critical temperature, it may have the lowest free energy and correspond to the ground state of the system. 

As discussed in section 3, the  zero temperature problem is defined by fixing the UV cutoff $\Lambda_{UV}$ and $\kappa_\infty$ \kappainfinity. In the BKT limit one can equivalently fix $\bar\mu=f(0)$ \dynfm\ and the ratio $C_1/C_2$ in the asymptotic solution \solintreg.

Turning on finite temperature modifies the full solution, but since the UV boundary condition remains the same, the asymptotic form of the solution must still take the form\foot{For the rest of this section we specialize to $d=3$.}  \solintreg,
\eqn\asttt{f(\rho)\sim \rho^{-{1 \over 2}} \left(C_1 \ln{\rho\over \bar\mu} +C_2 \right),}
with the same ratio of $C_1/C_2$.

It is convenient to measure energies in units of the horizon radius $r_h=\pi T$, and rewrite eq. \solintreg\ in the form
\eqn\solintrega{\eqalign{   f(\rho)|_{\kappa=0} \sim &r_h \left({r_h\over\rho}\right)^{1 \over 2}   
\left( C_1 \ln\left({\rho\over r_h}\right) +  C_1 \ln\left({r_h\over\bar\mu}\right)+C_2 \right)\cr
                                 =& r_h \left({r_h\over\rho}\right)^{1 \over 2}   
                                 \left(  C_1^{(T)} \ln\left({\rho\over r_h}\right) +C_2^{(T)} \right).\cr}  }
The constants $C_i^{(T)}$ defined in \solintrega\ satisfy the constraint 
\eqn\muTeq{  {\bar\mu\over r_h} = \exp\left[ {C_2 \over C_1 }-{C_2^{(T)}\over C_1^{(T)} }  \right].}
To study the finite temperature problem numerically, one can proceed as follows. 
Integrate the full equations of motion, starting with the initial conditions $f(0)=\bar\mu_T$, $f'(0)=0$.
The resulting solution asymptotes  to \solintrega\ for $\rho\gg r_h$. One can then read off $C_1^{(T)}, C_2^{(T)}$ and use \muTeq\ to calculate $\bar\mu/r_h$. Finally, one can invert the relation, and determine the $\bar\mu_T$ corresponding to the $\bar\mu$ of the zero temperature problem. 

\ifig\loc{$\bar\mu/r_h$ as a function of $\bar\mu_T/r_h$.}
{\epsfxsize3.3in\epsfbox{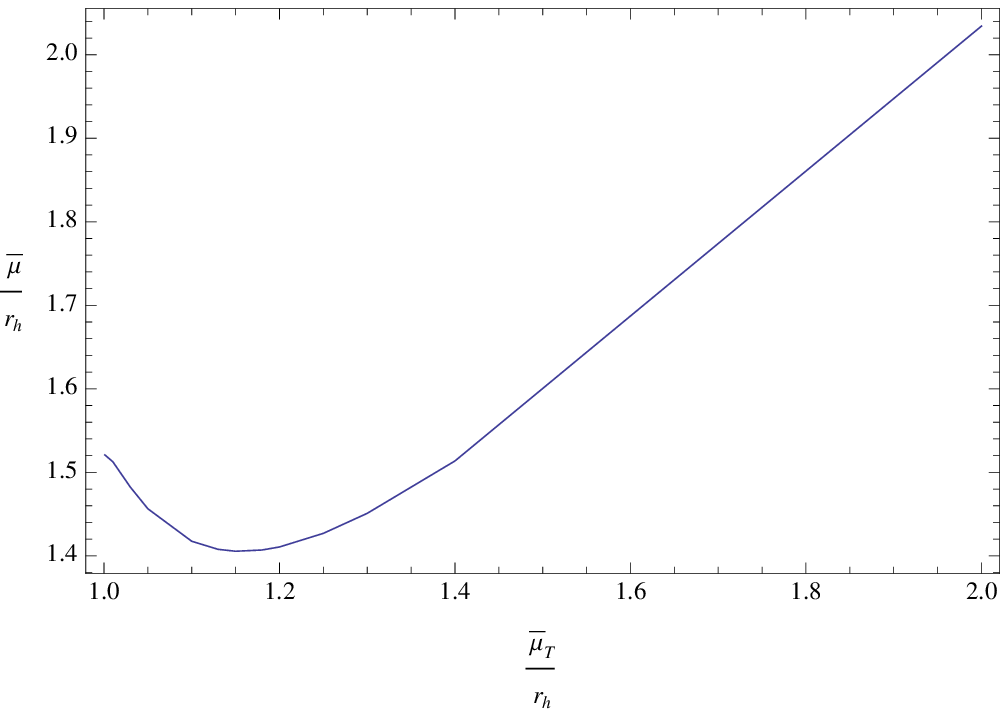}}

The resulting graph of $\bar\mu/r_h$ as a function of $\bar\mu_T/r_h$ is shown in fig. 5. At large  $\bar\mu_T/r_h$, it asymptotes to a straight line with a unit slope, as one would expect, since in this regime the effect of temperature is negligible and $\bar\mu\approx \bar\mu_T$. As the temperature increases, one encounters some non-trivial effects. One is that $\bar\mu/r_h$ is bounded from below. This implies that the Minkowski embedding does not exists above a certain critical temperature. Another is that there is a small region in which for a given $\bar\mu/r_h$ there are two distinct solutions with different values of $\bar\mu_T$ and different free energies.

\ifig\loc{$\bar\mu/r_h$ as a function of $\rho_0/r_h$.}
{\epsfxsize3.3in\epsfbox{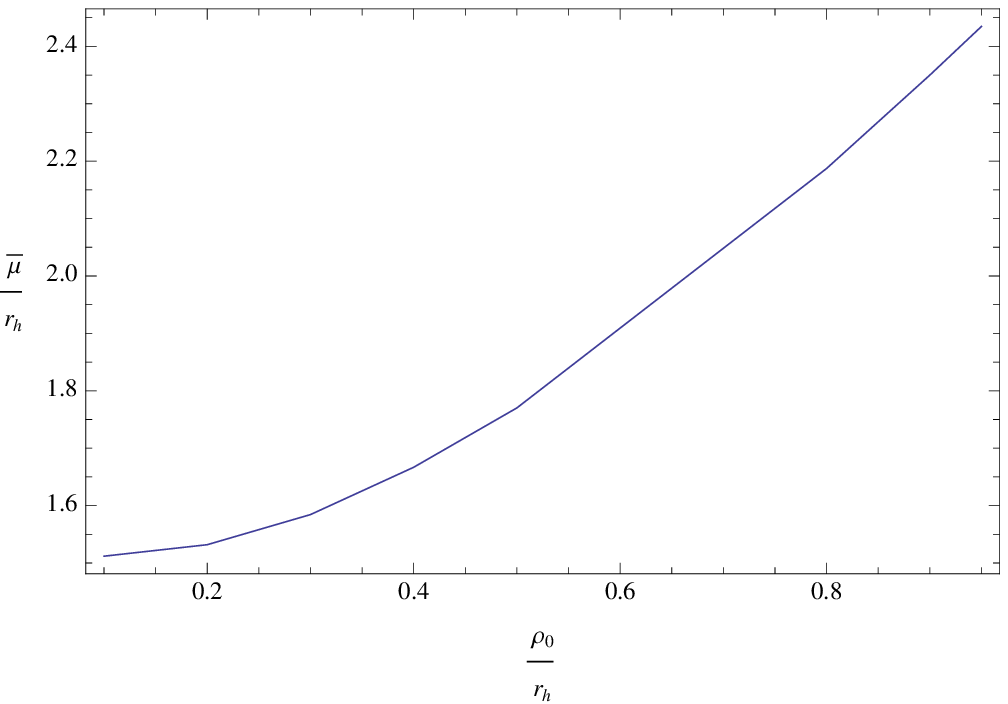}}

We now move on to embeddings of the probe brane which intersect the black hole horizon. Such embeddings can be parameterized by the position of the intersection in the $(\rho,x^9)$ plane,
$(\rho_0,\sqrt{r_h^2-\rho_0^2)}$. The relation between $\bar\mu/r_h$ and $\rho_0/r_h$ can again be determined numerically. The results, which are plotted in fig. 6, exhibit the following properties. As $\rho_0\ra 0$, the black hole embedding asymptotes to the limiting Minkowski embedding of figure 5, with $\bar\mu_T=r_h$. As $\rho_0\ra r_h$, the solution approaches the symmetric one, $x^9(\rho)=0$. It is also clear from figure 6 that symmetry breaking black hole embeddings only exist in a finite interval of temperatures, $r_h\in(1.51\bar\mu, 2.43\bar\mu)$.

To determine the phase structure of the theory, we need to compare the free energies \ftaction\ of the different solutions described above. As mentioned above, these free energies are UV divergent, but the differences are finite. It is convenient to subtract from the free energy that corresponding to the symmetric solution $f(\rho)=0$. The resulting free energies are shown in fig. 7.  

\ifig\loc{$\delta S$ as a function of $r_h/\bar\mu$ in the BKT limit. The blue curve corresponds to the Minkowski embeddings 
and the pink curve to the black hole embeddings. }
{\epsfxsize3.3in\epsfbox{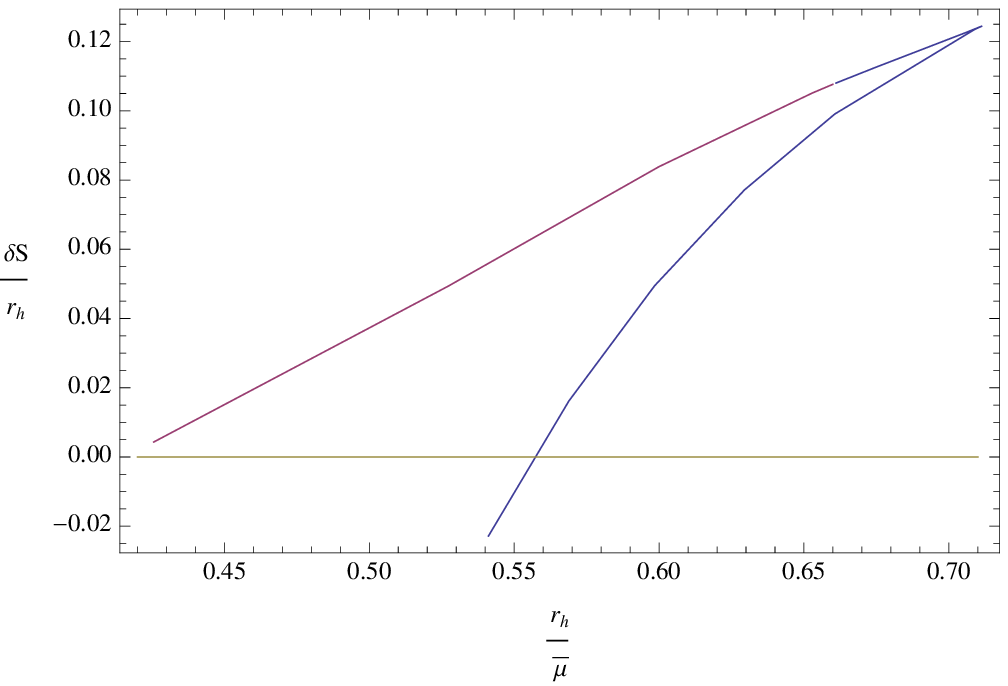}}

The blue curve corresponds to the Minkowski solution. At low temperatures, this (unique) solution is thermodynamically preferred, but for $T>T_c\approx 0.56\bar\mu/\pi$,
there is a first order phase transition to the symmetric solution, $f(0)=0$.
As the temperature is increased above $T=T_1 \approx  0.65 \bar\mu/\pi$, a second Minkowski solution appears, which merges with the first one at $T=T_2 \approx  0.71\bar\mu/\pi$.
Only the symmetric solution exists for $T>T_2$. In the interval $T\in(T_*,T_1)$ where $T_*\approx 0.42\bar\mu/\pi$, there exists another type of solutions, which originates from the black hole.
It is never thermodynamically preferred and represents a saddle point of the free energy.
It merges with the unstable branch of the Minkowski embeddings at $T=T_1$.

\subsec{Finite chemical potential}

Let us consider the situation where the temperature vanishes, but there is a non-vanishing
chemical potential conjugate to the fermion number.
In the dual description, the chemical potential  corresponds to the asymptotic value of the temporal component of the gauge field on the probe brane. 
Turning it on gives rise to a modification of the DBI action:
\eqn\fmuaction{S = \int d\rho\ r^{3-n}\rho^{n-1}\sqrt{1+f'^2-{(2\pi)^2\over L^4} A_0'^2}.}
The equation of motion for $\bar A_0= {(2\pi)\over L^2} A_0$ can be integrated to
\eqn\aoeom{   {r^{3-n}\rho^{n-1} \bar A_0'\over \sqrt{1+f'^2-\bar A_0'^2} }={\tilde d}^2.}
The conserved quantity ${\tilde d}$ is proportional to the charge density in the boundary theory. 
In order to have a non-vanishing ${\tilde d}$ (and non-vanishing flux on the worldvolume), the probe brane must intersect the horizon at $r=0$ which serves as a source of electric flux on the brane.
An important example of a solution with flux is given by the straight embedding  $f(\rho)=0$. Of course, it is still possible to have a Minkowski embedding where the $D$-brane does not intersect the horizon.
In this case the electric field on the worldvolume vanishes and $A_0=\mu_{ch}$ everywhere.

One can write $\mu_{ch} = A_0(\rho\ra\infty)$ as an integral
\eqn\muchem{  \mu_{ch} = {L^2\over 2\pi} \bar\mu_{ch}= {L^2\over 2\pi} \int_0^\infty d\rho \bar A_0', }
where the boundary condition $A_0(r=0)=0$ is ensured by requiring the temporal Wilson loop to be well defined at the point where the Euclidean time circle shrinks to zero size. This also ensures that at vanishing $\mu_{ch}$ the charge density vanishes. One can now relate $\tilde d$ to the chemical  potential $\mu_{ch}$. For the straight D-brane embedding, $f(\rho)=0$, one can solve for $A_0'$ using \aoeom\  and then integrate  \muchem\ to obtain  $\bar\mu_{ch}^{(f=0)} = (4 \Gamma(5/4)^2/\sqrt{\pi}) {\tilde d}$.

\ifig\loc{$\delta S$ as a function of $\bar\mu_{ch}/\bar\mu_0$: straight pink line - Minkowski embedding with $A_0=\mu_{ch}$; blue curve - $f(\rho)=0$ embedding; yellow curve - solution with flux which interpolates between the two.}
{\epsfxsize3.3in\epsfbox{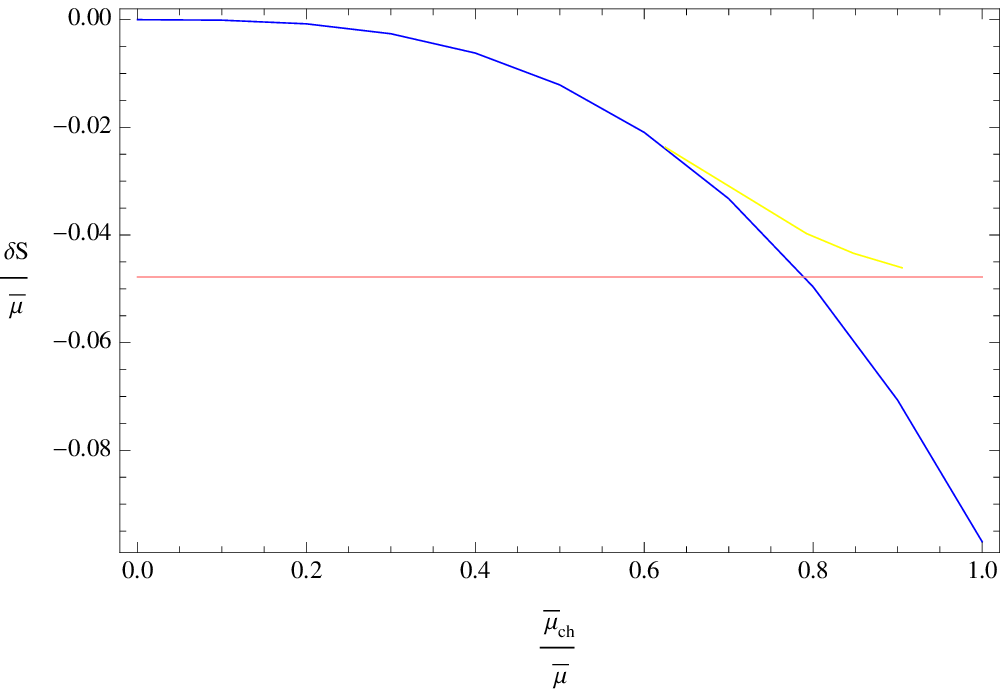}}

To obtain the phase diagram, let us first consider the Minkowski embedding and the straight $f(\rho)=0$ embedding with flux. We will fix $\bar\mu=f(0)$ for the Minskowski embedding and will measure
energies in units of $\bar\mu$ (note that according to \muchem, the natural scale for $\mu_{ch}$ is $\mu = \sqrt{\lambda} \bar\mu$, the scale of the fermion mass). The action of the Minkowski embedding is constant, $S_M=-0.0478 \bar\mu^3$. For the straight $D$-brane embedding, one can use \fmuaction\ to show that $S_{f=0} = -0.097 \bar\mu_{ch}^3$. Clearly, there is a first order phase transition between the Minkowski phase with vanishing density and the phase where the $D$-brane is straight and has a finite density. The complete phase diagram, shown in fig. 8, is a little more involved, due to the existence of a third type of solutions which have non-vanishing worldvolume flux and intersect the horizon, but have a nontrivial profile $f(\rho)$; they are discussed in Appendix B.

We have not analyzed the phase diagram with both temperature and chemical potential non-vanishing.
However, as in other D-brane systems, we expect a smooth line of first order phase transitions
in the $\mu_{ch}-T$ plane. This line presumably intersects the $\mu=0$ line at $T_c \sim\bar\mu$ and the $T=0$ line at $\mu_{ch,c}\sim \mu$, as discussed above.

\newsec{QCD}

Four dimensional QCD is expected to undergo a conformal phase transition as one varies the number of colors and flavors. The dynamics in the vicinity of this transition may lead to a phenomenologically viable model of dynamical electroweak symmetry breaking, known as ``walking technicolor'' (see \eg\ \HillAP\ for a review).  Since the transition occurs in a regime where the theory is strongly coupled, it is difficult to analyze its dynamics there using conventional field theoretic methods. One may hope that holographic techniques may shed light on it. In this section, we briefly review the expected structure near the phase transition from the point of view of our analysis. 

We start with a gauge theory with gauge group $G=SU(N)$ and $F$ flavors of Dirac fermions in the fundamental representation of the gauge group (see \eg\ \PeskinEV\ for a review). This theory is infrared free for $F\ge11N/2$; for smaller $F$ it is asymptotically free in the UV. As we review next, its IR dynamics depends on the number of flavors and colors.

If $F$ is only slightly smaller than $11N/2$, the gauge coupling runs from zero in the UV to a small non-zero value in the IR \BanksNN. The theory dynamically generates a scale, $\Lambda_{QCD}$, which can be thought of as the crossover scale between the UV and IR. For energies well above this scale, the correlation functions are dominated by their (asymptotically free) UV forms, while for $E\ll\Lambda_{QCD}$ they are dominated by the interacting IR fixed point. 

It is important to stress that the IR fixed point is an isolated CFT, which does not depend on any continuous parameters, unlike the defect theories studied in previous sections, which depended on the continuous parameter $\lambda$, the 't Hooft coupling of $\NN=4$ SYM.  Instead, the QCD fixed point depends on the discrete parameter $x=N/F$. We will treat this parameter as continuous by taking $N$ and $F$ to be large. 

As $F$ decreases (or $x$ increases), the IR theory becomes more strongly interacting. In particular, the dimensions of operators deviate more  and more from their free values. One can think of $x$ as a measure of the strength of the interactions in the IR CFT. Thus, it plays a role similar to that of $\lambda$ in the defect CFT discussed in previous sections. 

When the coupling exceeds a critical  value $x=x_c$, the model  undergoes a phase transition. For $x>x_c$, some of the symmetry of the IR CFT is dynamically broken. In particular, the $SU(F)_L\times SU(F)_R$ global symmetry, which is a symmetry of the Lagrangian preserved by the quantum dynamics for $x<x_c$, is spontaneously broken to the diagonal $SU(F)$. The conformal symmetry of the IR CFT is broken as well, at the scale $\mu\simeq \Lambda_{QCD}\exp(-a/\sqrt{x-x_c})$ mentioned in the introduction (see fig. 9). 

\ifig\loc{Conformal symmetry breaking scale $\mu$ as a function of the number of flavors $F$.}
{\epsfxsize3.3in\epsfbox{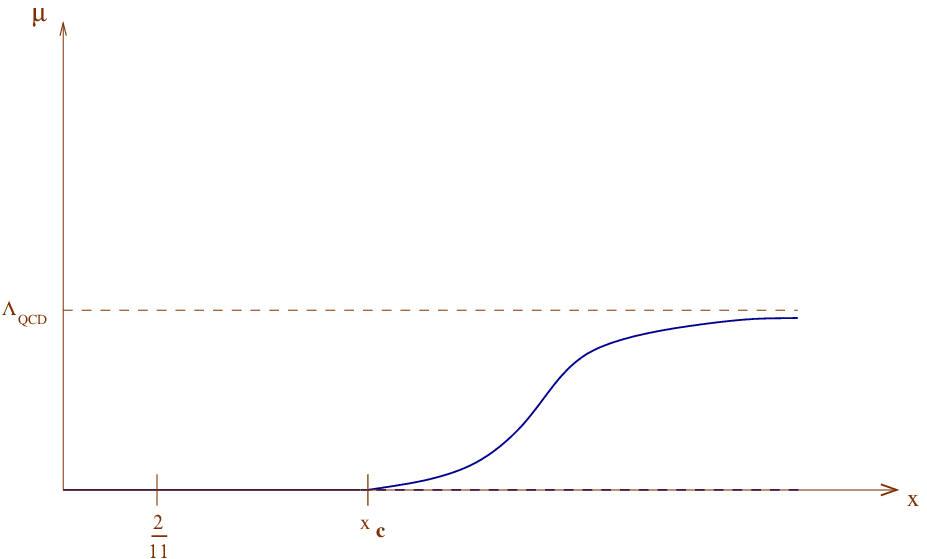}}

This transition is expected to be of BKT type. One can probe it by studying the vacuum expectation value of the operator $OW$ \ooww. For $x<x_c$, the IR theory is conformal, and this vev takes the form \oneptow. For $x>x_c$, the IR conformal symmetry is violated by a mechanism similar to that of section 2. As $x\to x_c$, the anomalous dimension $\gamma$ of $\bar\psi\psi$ approaches $-(d-2)/2=-1$ \CohenSQ, and beyond that point it formally becomes complex. The interpretation is the same as in the above discussion of the defect theory. At $x=x_c$, two fixed points of the RG which differ by the coefficient of the quartic double trace operator $(\bar\psi\psi)^2$ merge and annihilate (or move off to the complex plane). This generally leads to Miransky scaling of the dynamically generated scale  \KaplanKR. 

In the analysis of the earlier sections, we had a  conformal field theory for $\lambda<\lambda_c$. The analog of that here is the infrared fixed point of QCD for $x<x_c$. The RG flow of the gauge theory from its free UV fixed point to that IR theory corresponds from this point of view to a particular choice of UV cutoff. Hence $\Lambda_{QCD}$ corresponds in our previous discussion to the UV cutoff $\Lambda_{UV}$ \cutoffmap. The strength of the interactions of the fermions at the fixed point, which in our previous discussion was controlled by the 't Hooft coupling $\lambda$, is in QCD controlled by the parameter $x$ (\ie\ the number of colors and flavors) rather than the (running) coupling. To focus on the physics of the phase transition one can consider the double scaling limit $x\to x_c$, $\Lambda_{QCD}\to\infty$, with the dynamically generated scale $\mu$ held fixed. This is the analog of the BKT limit discussed in previous sections.

It is interesting to ask what our analysis of the CPT in defect theories teaches us about the corresponding phase transition in QCD. One question concerns the nature of the second fixed point that merges with the IR fixed point of QCD at $x=x_c$. For $x$ slightly below $x_c$, this fixed point can in principle be obtained by adding the (slightly irrelevant) double trace operator  $(\bar\psi\psi)^2$ to the Lagrangian, and following the resulting RG trajectory to its UV fixed point. Of course, in practice this is hard since the theory one perturbs is strongly coupled.

As $x$ decreases, the distance in coupling space between the two fixed points increases. An interesting question is whether the second fixed point exists all the way to weak coupling. In our analysis of defect theories, this was the case: setting $\lambda=0$ and adding the irrelevant four Fermi perturbation led to a non-trivial UV fixed point at which the dimension of the fermion bilinear was equal to one \anomdimuv, in agreement with \RosensteinNM.   

Thus, it is natural to conjecture that in four dimensional QCD the second fixed point also persists to arbitrarily weak coupling. There is a small subtlety here: the dimension of the fermion bilinear at the second fixed point is given by $d-1+\gamma_-$ \anomdimuv. As the gauge theory coupling approaches zero, which in QCD happens when $x\to 2/11$, this dimension approaches one. So at $x=2/11$, the operator $\bar\psi\psi$ becomes a decoupled free field, and adding $(\bar\psi\psi)^2$ to the Lagrangian simply gives mass to this field. In particular, it does not take one back to the free field theory one started with. This problem is related to the fact that the NJL model is not renormalizabe in $d=4$, but rather has a logarithmic residual dependence on the UV cutoff.

The above difficulty is avoided for any non-zero coupling, since then the analog of $C(\lambda)$ in \anomdimuv\ is strictly positive and the dimension of the fermion bilinear is larger than one. Thus, in QCD it is natural to conjecture that there is a new fixed point obtained by adding the four Fermi perturbation to the Lagrangian for any non-zero value of the gauge coupling.

In the defect problem, that fixed point can be analyzed via standard field theoretic techniques, described \eg\ in \RosensteinNM. It is interesting to ask whether one can establish its existence in four dimensional QCD as well. There are two complication in doing this. One is that as we just mentioned, one needs to turn on a non-zero gauge coupling. This is not a major problem since the fixed point should exist for arbitrarily small gauge coupling, and therefore should be accessible to standard weakly coupled field theory techniques. 

A more serious issue is that the fermions $\psi$ in QCD are large $N$ {\it matrices}, while in the defect problem they are {\it vectors}. More precisely, in both cases they are matrices of size $N\times F$, but while in the defect problem the transition takes place when $N$ is large while $F$ is of order one, in QCD it happens at a finite value of $x=N/F$. As is well known, while vector models are typically solvable at large $N$ using the methods reviewed in \RosensteinNM, matrix models are in general much more difficult to solve. Thus, the study of weakly coupled gauge theory with a four Fermi interaction is an open problem. It is natural to expect that this model does exhibit a non-trivial UV fixed point in which the dimension of the fermion bilinear is given by \anomdimuv. 

A widely discussed (and controversial) issue in the context of the CPT in QCD concerns the mass of the dilaton. The question is whether in the BKT limit, the mass of the lightest massive scalar goes to zero or remains comparable to the masses of the other mesons. This is an old debate; some recent contributions appeared in \refs{\AppelquistGY,\HashimotoNW}. 

It turns out that this issue is directly related to the vev of the trace of the stress tensor $T_\mu^\mu$. 
The authors of \AppelquistGY\ used the fact that  $\langle T^\mu_\mu\rangle\sim \beta(h)\langle\OO\rangle$, where $\beta(h)$ is the beta function for the coupling $h$ of an operator $\OO$, which in this case is a four fermi operator. They argued that the natural size of the vev of $\OO$ is $\mu^4$, which leads to $\langle T^\mu_\mu\rangle\sim \beta(h)\mu^4$. Since the beta function goes to zero in the BKT limit $x\to x_c$ they concluded that the expectation vaue of $T$ is parametrically suppressed relative to the natural mass scale of the theory, $\mu$. One can then show that the mass of the dilaton is suppressed as well. The authors of \HashimotoNW\ argued that in fact the expectation value of the stress tensor takes its natural size, $\langle T^\mu_\mu\rangle\sim \mu^4$, and the mass of the dilaton is correspondingly not suppressed. 

We calculated both the expectation value of the stress tensor and mass of the low lying $\sigma$ mesons for the defect theories. We found that the stress tensor \vacaction\ and mass of the lightest scalar meson \dilaton\ are not parametrically suppressed, in agreement with \HashimotoNW. As we mentioned in section 4, the latter does exhibit a mild finite suppression, which may also occur in the BKT regime of QCD and play a role in walking technicolor. In particular, the authors of \HashimotoNW\ found that the mass of the lightest scalar is smaller than that of the lightest vector meson by a factor of about 2.8. In our system this ratio is  (see \dilaton, \vectormass) about 2.6. Of course, these numbers cannot be compared directly, since the analysis of \HashimotoNW\ is not done in a controlled approximation, and ours is done in a different system. However, it is intriguing that the two calculations yield similar suppressions. 

Finally, we would like to emphasize that the conformal nature of the phase transition at $x=x_c$ in QCD is only a scenario, which has not been proven to date. It assumes that the transition has a sensible limit as $\Lambda_{QCD}\to\infty$, \ie\ it is continuous. An alternative scenario is that the system undergoes a first order phase transition  in which the dynamically generated scale jumps from zero to a scale of order $\Lambda_{QCD}$ at a value of $x$ for which the dimension of $\bar\psi\psi$ is larger than two. Of course, in that case the discussion of this section is inapplicable.

\newsec{Discussion}

In this paper, we showed that the system of defect fermions coupled to $\NN=4$ SYM \qftaction\ undergoes a phase transition between a weak coupling phase, in which the theory is scale invariant, and a phase in which the fermions acquire a dynamical mass. The order parameter exhibits Miransky (or BKT) scaling \bktscaling, a behavior also expected to hold in QCD near a critical number of flavors. 

At first sight, it is surprising that this type of continuous, infinite order transition is stable to large changes in the parameters of the theory. In other examples of systems with holographic duals, even when a transition is continuous at weak coupling, it typically becomes strongly first order when one goes to strong coupling. It seems that what protects the conformal phase transition in our system is its topological nature. It provides a realization of the picture proposed in \KaplanKR, according to which CPT's occur when two fixed points of the RG approach each other and annihilate as one changes the parameters of the model. This mechanism is robust -- as long as the physics varies continuously with the parameters, a transition that is conformal in one regime will remain so. 

In the class of systems \qftactiontwo, we saw this mechanism in action. By changing the parameters $d$ and $n$, we were able to push the transition to weak or strong coupling, but since the physics is expected to vary smoothly with these parameters, it is natural that it was found to be conformal in both regimes.  

Since the theory develops an arbitrarily small mass scale near the transition, it is interesting to ask whether the physics in this regime is universal. We provided descriptions of that physics at weak and strong coupling. In field theory, we discussed the order parameter $M(p)$ \momsp, \defm; its analog in  gravity was $f(\rho)$ \dsevenmetric, \dbia. The two order parameters satisfy very similar second order linear differential equations \cleggend,  \glrho, and encode the properties of the fermion bilinear near the transition at weak and strong coupling. An important difference between the two is that while in the strong coupling (gravity) regime we provided a precise definition of $f(\rho)$, as a scalar field describing the position of a probe brane in the bulk, in the field theory analysis we only defined $M(p)$  as the field measuring the deviation from conformality to leading order, \momsp.  A natural question is whether one can define $M(p)$  beyond the leading order and for all values of the coupling, such that for strong coupling it approaches $f(\rho)$. 

As a step in this direction, we would like to offer the following observation. Consider the action  
\eqn\genaction{S_{Dp} = \int d^dx \int d\rho \left({\rho^2\over\rho^2+f^2}\right)^{\half C(\lambda)}\rho^{d-1}\sqrt{1+f'^2}.}
 This action was obtained by replacing $n-d$ in \gaction\ by $C(\lambda)$ (and omitting an unimportant overall constant). The relation \cinfinity\ guarantees that \genaction\ gives the correct action for all $d$ and $n$ in the strong coupling limit $\lambda\to\infty$. However, there is no a priori reason for it to be reliable away from strong coupling. 
 
To see how well it does, it is interesting to take the opposite limit, $\lambda\to 0$, where $C(\lambda)$ is small  \formcl\ and one can expand in it. The equation of motion of \genaction\ in that limit can be read off \geom\ with the replacement $n-d\to C$. To compare to the perturbative analysis of appendix A, we take $f(\rho)=m+O(\lambda)$, and omit terms of order $\lambda^2$ or higher. This leads to:
\eqn\geomweak{\frac{\partial}{\partial\rho}\left(\rho^{d-1}f'\right) + C(\lambda)\rho^{d-1}{m\over(\rho^2+m^2)} = 0.}
Remarkably, this agrees precisely with the weak coupling result  (A.15), which was obtained by evaluating one loop Feynman diagrams! If this agreement is not accidental, it might indicate that the action \genaction\ provides a description of the physics near the phase transition {\it for all values of $d$ and $n$}.  In particular, it may be valid for the original $D3-D7$ system discussed in section 2, when $C(\lambda)$ is close to $1/4$. 

Of course, the action \geomweak\ depends on $C(\lambda)$, which is related to the anomalous dimension of $\bar\psi\psi$, and is only known in the two limits $\lambda\to 0,\infty$. However, its precise form is less important than the fact that it passes through the critical value  $(d-2)^2/4$. The physics near the transition depends on $\kappa$  \kappagendn, and ultimately on $\mu$ \dyngen. 

Although the discussion above is suggestive, there is clearly much to understand regarding universality in this class of systems. In particular, it would be interesting to show that when the dynamically generated mass is small relative to the UV cutoff, the physics of the order parameter is given by the action \genaction\ without $\alpha'$ corrections. It would also be interesting to include in the discussion the gauge fields dual to flavor currents, and extend the definition of the order parameter $M(p)$ in field theory beyond leading order. 

Other issues worth exploring include: 
\item{(1)} We focused on the case of one flavor brane $(F=1)$. One qualitatively new effect that appears in the $D3-D7$ system for $F>1$ is the breaking of the $U(F)$ flavor symmetry to $U(F_1)\times U(F_2)$ with $F_1+F_2=F$. In the brane picture this corresponds to a state in which $F_1$ branes are deformed to $x^9>0$ and $F_2$ to $x^9<0$.   The spontaneous breaking of the symmetry leads to the appearance of Nambu-Goldstone bosons which parametrize the coset $U(F)/U(F_1)\times U(F_2)$. They can be studied as in \SakaiCN, by analyzing the relevant components of the gauge field $A_\rho$.
\item{(2)} The model we studied is closely related to $\NN=4$ SYM, about which much is known due to its integrability at large $N$ (see \eg\ \BeisertJR\ for a review). The defect fermions can be viewed as replacing closed spin chains by open ones, with particular non-supersymmetric boundary conditions. It might be possible to use the description in terms of open spin chains to say more about the model, and in particular determine the function $C(\lambda)$, and perhaps understand the role of the action \genaction\ for general $\lambda$.
\item{(3)} Our discussion took place at large $N$, and it might be interesting to consider $1/N$ corrections. Beyond the leading order many aspects of our discussion need to be reconsidered. In particular, it is not obvious that the model is still conformal even at arbitrarily small $\lambda$. The $\NN=4$ SYM coupling in the bulk of course does not run, but one can in principle have a non-zero $\beta$-function for the boundary coupling in \qftactiontwo\ due to loops of fermions. 
\item{(4)} It might be interesting to generalize the discussion to other backgrounds, such as linear dilaton spacetimes. In these spacetimes there is an analog of the BF bound, and they exhibit holography \refs{\AharonyUB,\GiveonZM}. In this context it is useful to mention two dimensional string theory (or $c=1$ matrix model) \GinspargIS\ and the $3+1$ dimensional LST of the conifold (see \eg\ \GiveonZM) which involve condensates of closed string fields living at the boundary between the stable and unstable regions. 
\item{(5)} In \ReyZZ\ it was suggested that the $D3-D7$ system \dimensions\ might provide a model for the interactions of electrons in graphene. It would be interesting to realize the conformal phase transition of this system in this or other condensed matter system. 
\item{(6)} It would also be interesting to extend the picture presented in this paper for interacting defect fermions to QCD near the critical number of flavors $F_c$.

\bigskip

\noindent{\bf Acknowledgements}: We thank O. Aharony, E. Kiritsis, M. Kulaxizi, D. Mateos, M. Porrati, S. J. Rey, G. Semenoff, P. Wiegmann and J. Zaanen for discussions. This work is supported in part by DOE grant DE-FG02-90ER40560, the BSF -- American-Israel Bi-National Science Foundation and a VIDI innovative research grant from the Netherlands Organisation for Scientific Research (NWO). DK and AP thank the Weizmann Institute and the University of Chicago, respectively, for hospitality during part of this work.

\appendix{A}{Weak Coupling Analysis}

In this appendix we calculate the one loop anomalous dimension $\gamma(\lambda)$ of $\psibar\psi$ in the gauge theory  \qftactiontwo, and derive a nonlinear generalization of the differential equation \cleggend\ for $M(p)$. 

Adding a fermion mass term to \qftactiontwo\ leads to the action
\eqn\qftactionthree{\SS = \SS_{N=4}+
\int d^dx(i\bar{\psi}\slash\partial\psi + g\bar{\psi}\slash A\psi + g\bar{\psi}\Gamma_M\phi^M\psi - m\psibar\psi).}
Here and below, the indices $(a,b)$ and $(M,N)$ run over the $d$ directions along the defect and $(6-n)$ transverse dimensions respectively. The Feynman rules on the defect are:
\eqn\feynmanrules{\eqalign{\rm{Fermion\,\, propagator}  &: \frac{i(\slash{q}+m)}{q^2-m^2} ,  \cr
\rm{Scalar\,\, propagator} &: D_{MN}^\phi = \int\frac{d^{4-d}q_\perp}{(2\pi)^{4-d}}\frac{i\delta_{MN}}{q^2-q_\perp^2} = -i \frac{\delta_{MN}}{(-q^2)^{\frac{d-2}{2}}}\frac{\Gamma(\frac{d-2}{2})}{2^{4-d}\pi^{(4-d)/2}}, \cr
\rm{Gauge\,\,field\,\, propagator}  &:  D_{ab}^\gamma = %\int\frac{d^{4-d}q_\perp}{(2\pi)^{4-d}}\frac{1}{q^2+q_\perp^2}\left(\delta_{\mu\nu} - (1-\xi)\frac{q_\mu q_\nu}{q^2+q_\perp^2}\right)  = 
- \frac i {2(-q^2)^{d/2}}(2q^2g_{ab} - (d-2)q_a q_b(1-\xi))\frac{\Gamma(\frac{d-2}{2})}{2^{4-d}\pi^{(4-d)/2}}, \cr
\rm{vertex} &: ig\gamma_a t^i, \quad ig\Gamma_M t^i,
}}
where the scalar and gauge propagators are obtained by integrating the four dimensional propagators over the $(4-d)$ dimensions transverse to the intersection,  $q_\perp$; $t^i$ are $SU(N)$ matrices.
As discussed in Section 3, $\Gamma_M$ and $\gamma_a$ can be thought of as $SO(d+5-n,1)$ Dirac matrices. In particular, $\{\Gamma, \gamma\}=0,$ and in mostly minus signature $\{\Gamma^M, \Gamma^N\} = 2\eta^{MN} = -2\delta^{MN}$.

The one-loop fermion self energy
\eqn\dimensiondn{\epsfxsize3.5in\epsfbox{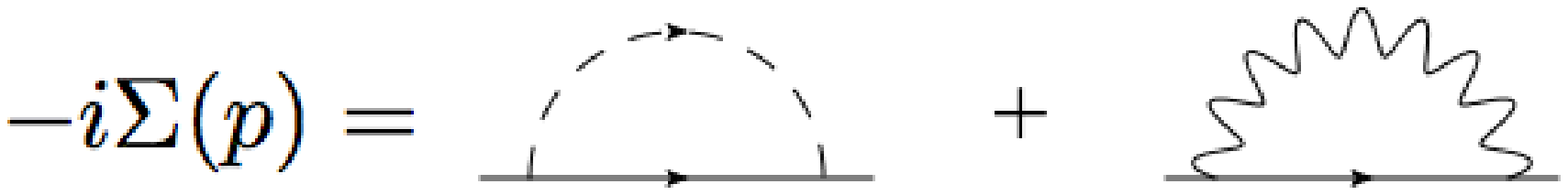}}
is given by
\eqn\sde{-i\Sigma(p) = -\frac{\lambda}{2}\int\frac{d^dk}{(2\pi)^d}
\left[D^\gamma_{\mu\nu}(p-k)\gamma^\mu\frac{\slash{k} + m}{k^2 - m^2}\gamma^\nu + D_{ab}^\phi(p-k)\Gamma^a\frac{\slash{k} + m}{k^2-m^2}\Gamma^b\right].}
Using the parametrization of $\Sigma$ in \formsig, one has:
\eqn\sdetwo{\eqalign{ B(p) &=  m - \frac {i\lambda} 2 \tr \int\frac{d^dk}{(2\pi)^d}\left[D_{\mu\nu}^\gamma(p-k)\gamma^\mu\frac{m}{k^2 - m^2}\gamma^\nu + D_{ab}^\phi(p-k)\Gamma^a\frac{m}{k^2 - m^2}\Gamma^b\right]
\cr
A(p) &= 1 + \frac{i\lambda} 2 \tr \int\frac{d^dk}{(2\pi)^d}\frac{\slash{p}}{p^2}\left[ D_{\mu\nu}^\gamma(p-k)\gamma^\mu \frac{\slash{k}}{k^2-m^2}\gamma^\nu +  D_{ab}^\phi(p-k)\Gamma^a \frac{\slash{k}}{k^2-m^2}\Gamma^b\right].
}}
After Wick rotating and inserting the propagator, one finds
\eqn\propagatortwo{\eqalign{ B(p) &=  m  + \frac{\lambda}{2} \frac{\Gamma(\frac{d-2}{2})}{16\pi^{(4+d)/2}} (d-\frac{d-2}{2}(1-\xi) + 6-n )\int d^dk\frac{1}{|p-k|^{d-2}}\frac{m}{k^2+m^2}
\cr
A(p) &= 1 + \frac{\lambda}{2} \frac{\Gamma(\frac{d-2}{2})}{16\pi^{(4+d)/2}} \int d^dk \frac{1}{p^2}\frac{1}{|p-k|^{d-2}}\frac{1}{k^2+m^2}
\cr 
&\times \tr \frac1 d\left[((d-2)+(6-n))\slash{p}\slash{k} + \frac{(d-2)(1-\xi)\slash{p}(\slash{p}-\slash{k})\slash{k}(\slash{p}-\slash{k})}{2(p-k)^2} \right],}}
the second of which can be rewritten as
\eqn\atwo{\eqalign{A(p) &= 1 +  \frac{\lambda}{2} \frac{\Gamma(\frac{d-2}{2})}{16\pi^{(4+d)/2}} \int d^dk \frac{(d-2)+(6-n)}{2(k^2+m^2)}\left[\frac{1}{|p-k|^{d-2}} + \frac{k^2}{p^2|p-k|^{d-2}} - \frac{1}{p^2|p-k|^{d-4}}\right]
\cr
&+\frac{(d-2)(1-\xi)}{4(k^2+m^2)}\left[-\frac{1}{|p-k|^{d-2}} - \frac{k^2}{p^2|p-k|^{d-2}} + \frac{k^4}{p^2|p-k|^d} + \frac{p^2}{|p-k|^d} - \frac{2k^2}{|p-k|^d} \right].}}
Now, recall the identity for the Green's function of the Laplacian:
\eqn\deltafunction{\nabla^2 \frac{1}{r^{d-2}} = -\frac{(d-2)2\pi^{d/2}}{\Gamma(\frac d 2)}\delta^{(d)}(r).}
Applying it to \propagatortwo, \atwo\ yields
\eqn\clegexacta{\eqalign{&\frac{\partial}{\partial p}(p^{d-1}B'(p)) + \frac \lambda 2 \frac{1}{4\pi^2}\left(d-\frac{d-2}{2}(1-\xi) + 6-n\right)p^{d-1} \frac{m}{p^2+m^2} = 0,
\cr
&\frac{\partial}{\partial p}(p^{d-1}A'(p)) + (d-2)\left((6-n+d-2)\frac{\lambda}{8\pi^2d} -\frac{(d-2)}{2}(1-\xi)\frac{\lambda}{8\pi^2(d-2)}\right)p^{d-1} \frac{1}{p^2+m^2} = 0.}}
To leading order in $m$, eqs. \clegexacta\ are solved by
\eqn\abdefs{\eqalign{B(p) &= m\left(1 + \beta\ln\left(\frac{p}{\mu}\right)\right) \simeq m\left(\frac{p}{\mu}\right)^\beta, 
\cr
A(p) &= 1 + \alpha\ln\left(\frac{p}{\mu}\right) \simeq \left(\frac{p}{\mu}\right)^\alpha,
}}
 where the coefficients are %
\eqn\abdef{\eqalign{\alpha &= -\frac{\lambda}{8\pi^2}\left((6-n+d-2)\frac{1}{d} -\frac{(d-2)}{2}(1-\xi)\frac{1}{d-2}\right) \cr
\beta &= -\frac{\lambda}{8(d-2)\pi^2}\left(d-\frac{d-2}{2}(1-\xi) + 6-n\right)
}}
and $\mu$ is the renormalization scale. Note that in the absence of the scalar fields (\ie\ setting $n=6$) and for $d=4$, $\alpha$ is proportional to the gauge parameter $\xi$, in agreement with \AppelquistWR. As another check, the one-loop values of $\alpha, \beta$ in four-dimensional Yukawa theory and QED are well-known results, derived for example in \PeskinEV. Our results agree with theirs under the map $g^2 \leftrightarrow \lambda/2$.

The quantities derived in \abdefs\ can be combined to form the gauge-invariant part of the fermion self-energy \defm:
\eqn\sigmahat{M(p)= \frac{B(p)}{A(p)} = m\left(\frac{p}{\mu}\right)^{\gamma},}
where
\eqn\anomdim{\gamma(\lambda) = \beta - \alpha = -\frac{\lambda}{4d(d-2)\pi^2}\left(2(d-1)+ (6-n) \right).}
$\gamma(\lambda)$ is the anomalous dimension of $\psibar\psi$, \fermbi. It is gauge invariant, as expected.

In section 3, we saw that $M(p)$ satisfies eq. \cleggend\ at large momentum. That equation, and its nonlinear generalization which extends to all $p$, can be derived to order $\lambda$ from \clegexacta\ in the following way. Expanding $A(p) \simeq 1 + A_1(p) + O(\lambda^2)$, $B(p) \simeq m(1 + B_1(p) + O(\lambda^2))$ and using \sigmahat, one finds
\eqn\mexpand{M(p) \simeq m(1 + B_1(p)- A_1(p)) + O(\lambda^2).}
If we now take the difference of the equations in \clegexacta, we arrive at
\eqn\mtwo{\frac{\partial}{\partial p}(p^{d-1}M'(p)) + C(\lambda)p^{d-1}\frac{m}{p^2+m^2} + O(\lambda^2) = 0,}
where\foot{Note that this is consistent with eq. \anomdimc\ when $C(\lambda)$ is small: there we can expand $$\gamma(\lambda) \simeq -\frac{d-2}{2} + \frac{d-2}{2}\left(1 - \frac 1 2\left(\frac{2}{d-2}\right)^2C(\lambda) \right) = -\frac{C(\lambda)}{d-2}.$$}
\eqn\clambdaone{C(\lambda) = -(d-2)\gamma(\lambda) = \frac{\lambda}{4d\pi^2}(2(d-1)+(6-n)) + O(\lambda^2).}
In particular, the $(d,n) = (2,6)$ result for $C(\lambda)$ agrees with \KaplanKR, and the $(d,n) = (3,5)$ result for the $D3/D7$ system of section 2 is
$$C(\lambda) = \frac{5\lambda}{12\pi^2} + O(\lambda^2) .$$
Finally, we replace $m$ with $M(p)$, which is consistent at leading order in $\lambda$, and find:
\eqn\mthree{\frac{\partial}{\partial p}(p^{d-1}M'(p)) + C(\lambda)p^{d-1}\frac{M(p)}{p^2+M(p)^2} + O(\lambda^2) = 0.}
For large $p$ (or small $M$) this reduces to \cleggend.

Eq. \mthree\ describes the vicinity of the conformal phase transition at weak coupling. It is an improvement over \cleggend\ which breaks down at momenta on the order of the dynamically generated scale, and can be thought of as the analogue of the nonlinear strong coupling equation \geom\ following from the DBI action.

Near $d=2$, where weak coupling techniques are applicable, \mthree\ can be alternatively obtained by using the Schwinger-Dyson equations in the rainbow approximation. These are given diagrammatically by
\eqn\dimensiondn{\epsfxsize4in\epsfbox{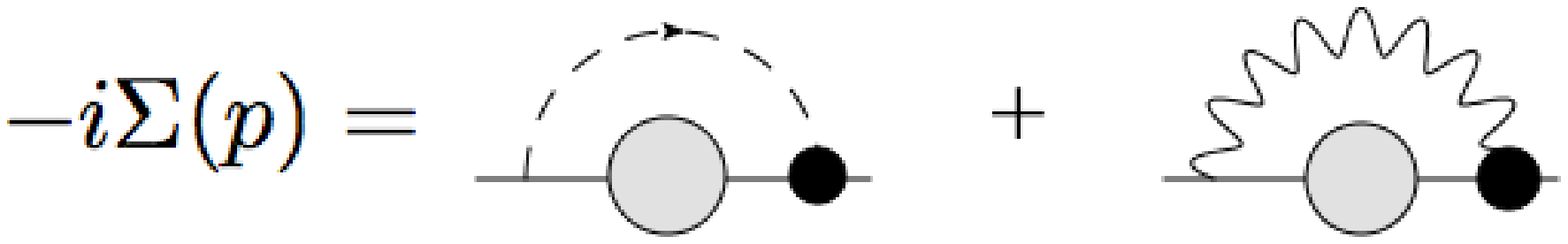}}
where the lines with grey dots are full (as opposed to bare) propagators, and the black dots denote the full vertices $\Delta_\phi^a = ig\Gamma^a + O(g^3), \Delta_\gamma^\mu = ig\gamma^\mu + O(g^3)$. Then Eq. \sde\ is promoted to
\eqn\sdetwo{-i\Sigma(p) = -\frac{\lambda}{2}\int\frac{d^dk}{(2\pi)^d}
\left[D^\gamma_{\mu\nu}(p-k)\gamma^\mu\frac{A(k)\slash{k} + B(k)}{A(k)^2k^2 - B(k)^2}\Delta_\gamma^\nu + D_{ab}^\phi(p-k)\Gamma^a\frac{A(k)\slash{k} + B(k)}{A(k)^2k^2-B(k)^2}\Delta_\phi^b\right]}
which, again, can be separated into different spinor structures:
\eqn\sdethree{\eqalign{ B(p) &=  m - \frac {i\lambda} 2 \tr \int\frac{d^dk}{(2\pi)^d}\left[D_{\mu\nu}^\gamma(p-k)\gamma^\mu\frac{B(k)}{A(k)^2k^2 - B(k)^2} \Delta_\gamma^\nu + D_{ab}^\phi(p-k)\Gamma^a\frac{B(k)}{A(k)^2k^2 - B(k)^2}\Delta_\phi^b\right]
\cr
A(p) &= 1 + \frac{i\lambda} 2 \tr \int\frac{d^dk}{(2\pi)^d}\frac{\slash{p}}{p^2}\left[ D_{\mu\nu}^\gamma(p-k)\gamma^\mu \frac{A\slash{k}}{A(k)^2k^2-B(k)^2}\Delta_\gamma^\nu +  D_{ab}^\phi(p-k)\Gamma^a \frac{A\slash{k}}{A(k)^2k^2-B(k)^2}\Delta_\phi^b\right].
}}
Now by eq. \abdef, the deviation of $A(p)$ from unity is suppressed by $(d-2)$ relative to $B(p)$. It is therefore consistent near $d=2$ to set $A(p) =1$, and replace $B(p)$ with $M(p)$. Likewise, the gauge-dependent part of the gauge field propagator \feynmanrules\ is suppressed by $(d-2)$, so we are free to fix $\xi=1$ in the following. Finally, to order $\lambda$, it is a consistent truncation to work in the quenched approximation and replace the full vertices with bare ones.

With these approximations and after Wick rotating, the first line of \sdethree\ becomes
\eqn\propagatorthree{ M(p) =  m  + \frac{\lambda}{2} \frac{\Gamma(\frac{d-2}{2})}{16\pi^{(4+d)/2}} (d+ 6-n )\int d^dk\frac{1}{|p-k|^{d-2}}\frac{M(k)}{k^2+M(k)^2}.}
Upon applying the Laplacian, we arrive at the differential equation
\eqn\clegexactthree{\frac{\partial}{\partial p}(p^{d-1}M'(p)) + \frac \lambda 2 \frac{1}{4\pi^2}\left(d + 6-n\right)p^{d-1} \frac{M(p)}{p^2+M(p)^2} = 0.}
The deviation of \clegexactthree\ from \mthree\ appears at order $(d-2)$, and presumably is reconciled when one takes higher order corrections into account.

\appendix{B}{Some holographic results in the BKT limit}

\subsec{Action}

Let us investigate the behavior of the action \gaction\ in the BKT regime.
It is intuitively clear that in the limit $\kappa\ra0$ the action approaches a
constant whose scale is set by $\mu$.
We will see momentarily that this is indeed the case.
It is technically convenient to divide the region of integration in \gaction\
into the small $\rho$ part and the large $\rho$ part.
Let them be separated by $\rho_*$, such that $\mu\ll\rho_*\ll\Lambda_{UV}$,
so that the total action consists of two terms, 
\eqn\actot{ S[f]= S[f]_{0\le\rho\le\rho_*}+ S[f]_{\rho_*\le\rho\le\Lambda_{UV}}  . }
In the large $\rho$ region the action is approximately quadratic in $f(\rho),f'(\rho)$
and therefore one can use equations of motion to show that
the non-vanishing contribution to the integral only
comes from the boundary term at $\rho=\rho_*$:
\eqn\Sbndry{   S[f]_{\rho_*\le\rho\le\Lambda_{UV}} = -{1\over2} \left[\rho^2f'(\rho)\right]'f(\rho)|_{\rho=\rho_*}.  }
The boundary term coming from the UV cutoff vanishes precisely because $f(\rho=\Lambda_{UV})=0$.
The integral from $\rho=0$ to $\rho=\rho_*$ can be evaluated numerically.
The result of this numerical integration, $S[f]_{0\le\rho\le\rho_*}$, with the boundary term \Sbndry\ added,
converges to $-C \mu^3 V_3$, where $C$ is a positive numerical factor
and $V_3$ is the volume of the three-dimensional spacetime.
This implies, according to \dyngen, that
\eqn\actionbkt{  S_0\sim - \Lambda^3 \exp( -{3 \pi\over\sqrt{\kappa}})   }
consistent with the BKT-type phase transition.
Here the zero in the subscript indicates the vacuum nature of the solution.

\subsec{Chemical potential}

There are three types of embeddings: Minkowski embeddings with vanishing gauge flux on the woldvolume of the D-brane (so that $A_0=\mu_{ch}$), straight  embeddings $f(\rho)=0$ with flux, and curved embeddings with flux which start at $\rho=0$ and can be parameterized by the initial condition $f'(0)$. The first two were discussed in Section 5.2.

To analyze embeddings with flux and nontrivial profile, it is technically convenient to measure all energies in the problem in units of ${\tilde d}$. For a fixed value of ${\tilde d}$, one can vary $f'(0)$ to obtain the solutions with various asymptotics (in practice, one integrates the equation of motion for $f(\rho)$ which follows from \fmuaction\ while eq. \aoeom\ is used to eliminate $\bar A_0'$ in favor of ${\tilde d}$).
The value of $\bar\mu_{ch}/{\tilde d}$ can then be obtained by integrating \aoeom\ numerically.
Finally, the value of $\bar\mu/{\tilde d}$ for a given solution can also be obtained from the
asymptotic behavior, in a similar way as the values of $\bar\mu/r_h$ and $\bar\mu_T/r_h$ 
were obtained in the finite temperature case, where eq. \muTeq\ was used.
One can therefore compute the ratio $\bar\mu_{ch}/\bar\mu$ as a function of $f'(0)$.
This turns out to be a monotonically increasing function, which starts from $\bar\mu_{ch}/\bar\mu\approx 0.6$
(which corresponds to the $f'(0)\ra0$ limit, where the solution joins the straight $f=0$ solution with flux)
and approaches unity from below
(which corresponds to $f'(0)\ra\infty$, the regime where the curved solution with flux joins the corresponding
Minskowski solution).
The action can be computed numerically; the results are shown in Fig. 8.
These curved solutions with flux are never thermodynamically preferred.

\listrefs

\bye